\documentclass[%
 reprint,
 amsmath,amssymb,
 aps,
]{revtex4-2}
\usepackage{physics}
\usepackage[pdftex]{graphicx}
\usepackage{dcolumn}
\usepackage{bm}

\begin{document}

\title{Hydrodynamic synchronization and collective dynamics \\of colloidal particles driven along a circular path}

\author{Takumi Miyamoto}
\author{Masayuki Imai}%
\author{Nariya Uchida}%
 \email{uchida@cmpt.phys.tohoku.ac.jp}
\affiliation{%
Department of Physics, Tohoku University, Sendai 980-8578, Japan
}%

\date{\today}

\begin{abstract}
We study theoretically the collective dynamics of particles driven by an optical vortex
along a circular path.
Phase equations of $N$ particles are derived by taking into account both 
hydrodynamic and repulsive interactions between them.
For $N=2$, the particles attract with each other and synchronize, 
forming a doublet that moves faster than a singlet. 
For $N=3$ and $5$, we find periodic rearrangement of doublets and a singlet. 
For $N=4$ and $6$, the system exhibits either a periodic oscillating state or 
a stable synchronized state depending on the initial conditions.
These results reproduce main features of previous experimental findings.
We quantitatively discuss the mechanisms governing the non-trivial collective dynamics. 
\end{abstract}

\maketitle

\section{\label{sec:level1}Introduction}
Hydrodynamic synchronization is considered to be a key mechanism 
that controls the collective dynamics of flagella and cilia.
For example, bacteria such as {\it E. coli} shows run-and-tumble motion 
induced by dynamical bundling/unbundling of their flagella,
and ciliates such as {\it Paramecium} regulates the direction of motion 
using metachronal waves\cite{bray2001cell}.
Metachronal waves of cilia also play vital roles in moving fluid in mammalian bodies  
(airways, embryo, ventricles, etc.) 
These phenomena suggest the importance of indirect hydrodynamic interaction 
between flagella and cilia in their coordinate motion. 
Complexity due to dynamical regulation by molecular motors is  
removed by considering an artificial model system. 
Experiments have been pushed forward using colloidal particles driven by optical tweezers,
and verified theoretical scenarios in the emergence of synchronization~\cite{Maestro2018, Bruot-Cicuta, Brumley}: 
dynamical switching of the driving potential (rower model)\cite{Consentino, Bruot},
deformability of the trajectory~\cite{NiederMayer}, and 
periodic modulation of the driving force~\cite{Uchida2011,Uchida2012,Uchida2017, Kotar}.
The last mechanism enables flexible control of the phase difference between two rotors~\cite{Maestro2018}. 
Fluid inertia may also induce synchronization in a finite-Reynolds number system~\cite{Mario, Oyama2018}.


In this paper, we theoretically address the dynamics of colloidal particles driven along a circular path. 
Non-trivial collective motion have been found in experiments 
using an optical vortex~\cite{Roichman, Sokolov, Sassa, Okubo2015, Kimura2017, Nagar} 
or optical tweezers~\cite{Lutz, Kotar2010, Lhermerout_2012, Kavre, Simpson, Leonardo}.
An optical vortex can drive multiple particles simultaneously along a single circular trajectory~\cite{Curtis}, 
while optical tweezers manipulate the particles individually.
For three particles, a limit-cycle behavior predicted in Ref.~\cite{Reichert2004} 
is confirmed and effects of non-constant force profiles have been discussed~\cite{Lutz,Roichman}.
Sokolov {\it et al.}~\cite{Sokolov} showed that a pair of particles attract each other due to radial displacements 
from the guided circular trajectory. Increasing the particle number $N$ up to $N=11$,
they also found that the collective mobility changes non-monotonically: 
for odd $N$, one unpaired particle remains and reduces the collective mobility.  
Sassa {\it et al}.~\cite{Sassa} performed experiments for up to $N=9$ particles
and observed multiple collective states for even $N$:
the particles either form stable doublets (stable states) or 
exhibit periodic rearrangement of doublets and (a) singlet(s) (oscillating states),
and switching between these states are observed.
More complex dynamics are observed for particles of different sizes\cite{Okubo2015}
and particles in confined geometry \cite{Saito2018}.
Theoretically, linear stability analysis~\cite{Reichert2004,Sassa} and 
direct numerical simulations~\cite{Sokolov,Okubo2015} have been done 
to capture the essential mechanism of the collective motion.
In the present paper, we pursue an analytical approach to  
gain more quantitative description of the phenomena.
Employing the phase reduction method, which was successfully applied to 
hydrodynamic synchronization in a system with deformable trajectories~\cite{NiederMayer,Maestro2018},
we obtain the dynamical equations in a compact form.
Numerical solutions of the phase equations for up to $N=6$ reproduce
the main features of the experimental findings.
Construction of the paper is as follows:
In Section II, we describe the model and derive the phase equations.
In Section III, numerical solution of the phase equations are presented 
for $N=2 - 6$ particles.
In Section IV, we discuss the results in comparison with the experiments.

\section{\label{sec:level2} Model}
\subsection{\label{sec:level21} Geometry and Forces}
We consider $N$ colloidal particles of radius $a$ that are 
driven along a circular trajectory of radius $R$ on the $z=0$ plane.
The position of the $i$-th particle ($i=1,2,...,N$) is given by
\begin{equation}
\textbf{r}_i = R_i \textbf{e}_i^r = (R_i \cos{\phi_i},R_i \sin{\phi_i},0),
\label{position}
\end{equation}
where $R_i = R_i(t)$ is the instantaneous orbital radius of the $i$-th particle,
which can deviate from the target radius $R$ set by the optical vortex,
and $\phi_i = \phi_i(t)$ is the phase.
Here and hereafter we use the radial and tangential unit vectors
\begin{align}
\textbf{e}_i^r & =(\cos{\phi_i},\sin{\phi_i},0) 
\end{align}
and 
\begin{align}
\textbf{e}_i^\phi & = (-\sin{\phi_i},\cos{\phi_i},0).
\label{base}
\end{align}
We label the particles along the counter-clockwise direction so that $\phi_1 < \phi_2 < \cdots < \phi_N$,
and define the phase difference 
\begin{equation}
\Delta_{ij} = \phi_j - \phi_i.
\end{equation}
The motion of the particles are governed by three types of forces acting on them: 
the forces due to the optical vortex, 
the hydrodynamic forces, and 
the repulsive interaction among the particles.


The force by the optical vortex is decomposed into 
the radial and tangential components as 
\begin{equation}
\textbf{F}_i = F_i^r \textbf{e}_i^r + F_i^\phi \textbf{e}_i^\phi.
\label{force}
\end{equation}
We assume that the tangential driving force is constant,
\begin{equation}
F_i^\phi = F = {\rm const.},
\label{Fphi}
\end{equation}
while the radial restoring force is linearly proportional 
to the radial displacement:
\begin{equation}
F_i^r = -k_r \delta R_i, \qquad \delta R_i = R_i - R.
\label{harmonic}
\end{equation} 
The spring constant $k_r$ is determined by the curvature of the optical trapping potential
along the radial direction. 
We will use the dimensionless spring constant
\begin{equation}
\kappa = \frac{k_r R}{F}
\end{equation} 
to describe the relative strength of trapping.
We assume strong trapping ($\kappa \gg 1$) and regard $\kappa^{-1}$ as a small perturbative parameter.


The particles move in an incompressible fluid of shear viscosity $\eta$. 
Since we consider a micrometer-sized system, the Reynolds number is negligibly small 
(typically $Re \sim 10^{-5}$) 
and the flow velocity field ${\bf v}({\bf r})$ obeys the Stokes equation,
\begin{align}
\eta \nabla^2 \textbf{v} - \nabla p & = 0 , \\
\nabla \cdot \textbf{v} & = 0 .
\label{stokes}
\end{align}
The hydrodynamic drag force exerted on the $i$-th particle is given by
\begin{equation}
\textbf{g}_i = \gamma [\textbf{v}(\textbf{r}_i)-\dot{\textbf{r}}_i], 
\label{resistance}
\end{equation}
where $\gamma = 6\pi \eta a$ is the drag coefficient.
The flow velocity is determined by
\begin{equation}
\textbf{v}(\textbf{r}_i)\simeq -\sum_{j\neq i}\textbf{G}(\textbf{r}_i,\textbf{r}_j)\cdot \textbf{g}_j.
\label{two_body_approx.}
\end{equation}
Here, $\textbf{G}(\textbf{r}_i- \textbf{r}_j )$ 
is the off-diagonal elements of the grand mobility tensor
and describes the hydrodynamic interaction between the particles.
We use the Rotne-Prager-Yamakawa approximation~\cite{Rotne1969, Yamakawa},
\\
\\
\begin{multline}
\textbf{G}(\textbf{r}) = 
\frac{1}{8\pi\eta r} \left( \textbf{1} + \frac{\textbf{r} \textbf{r} }{r^2} \right) + 
\frac{a^2}{12\pi \eta r^3} \left( \textbf{1} - \frac{3 \textbf{r} \textbf{r} }{r^2} \right).
\label{Ossen}
\end{multline}
The first term on the RHS of (\ref{Ossen}) is the Oseen tensor, 
which is the Green function of the Stokes equation, 
and the second term is the correction due to finite volume of the spheres.
Introducing the dimensionless tensor
\begin{equation}
\hat{\bf G}_{ij} =  \gamma {\bf G}({\bf r}_i - {\bf r}_j)
\end{equation}
which scales as $a/|{\bf r}_i - {\bf r}_j|$, 
we find that 
\begin{equation}
\alpha = \frac{a}{R}
\end{equation}
is the dimensionless parameter describing the relative strength of 
the hydrodynamic interaction.
Substituting (\ref{two_body_approx.}) into (\ref{resistance}) 
and then back into (\ref{two_body_approx.}), we get an expression of 
the velocity field to the second order with respect to $\gamma {\bf G}$.
This can be continued recursively to give a series expansion of the velocity field 
in terms of $\alpha$. Neglecting $O(\alpha^2)$ terms, we get the linear relation
\begin{equation}
{\bf v}({\bf r}_i) \simeq \sum_{j\neq i} \hat{\bf{G}}_{ij} \cdot \dot{\textbf{r}}_j.
\label{vapprox}
\end{equation}


The intermolecular forces cause excluded volume interaction between the particles.
Assuming that the radial displacements of the particles are small compared to their radii, 
we can neglect the radial component of the repulsive force.
Thus the repulsive force exerted by the $j$-th particle on the $i$-th is written in the form 
\begin{align}
\textbf{F}_{ji}^{rep} &= F_{ji}^{rep} \textbf{e}_i^{\phi}, &
\\
F_{ji}^{rep} &= \frac{1}{R}\frac{\partial U(\Delta_{ij})}{\partial \Delta_{ij}}, &
\label{repulsive}
\end{align}
where $U(\Delta)$ is the potential as the function of the phase difference.

We choose the function~\cite{Reichert2004}
\begin{equation}
U(\Delta) = U_0 
\qty(\frac{R |\Delta|}{2 a} - 1 )^{-12},
\label{repU}
\end{equation}
where $U_0$ is a positive constant, and $\Delta$ is the phase difference folded into $[-\pi,\pi]$.
Because this potential diverges when the particles are in contact $(\Delta=2a/R)$,
we can avoid overlapping of particles in numerical analysis.
Since the force decays rapidly as a function of the distance, 
we retain only the interaction 
between the nearest neighbor pairs ($i$ and $i \pm 1$ mod $N$). 
The relative strength of the repulsive force 
compared to the driving force is given by the dimensionless parameter
\begin{equation}
u = \frac{U_0}{Fa}.
\end{equation}

\begin{figure}[h]
\includegraphics[clip, width=8.0cm]{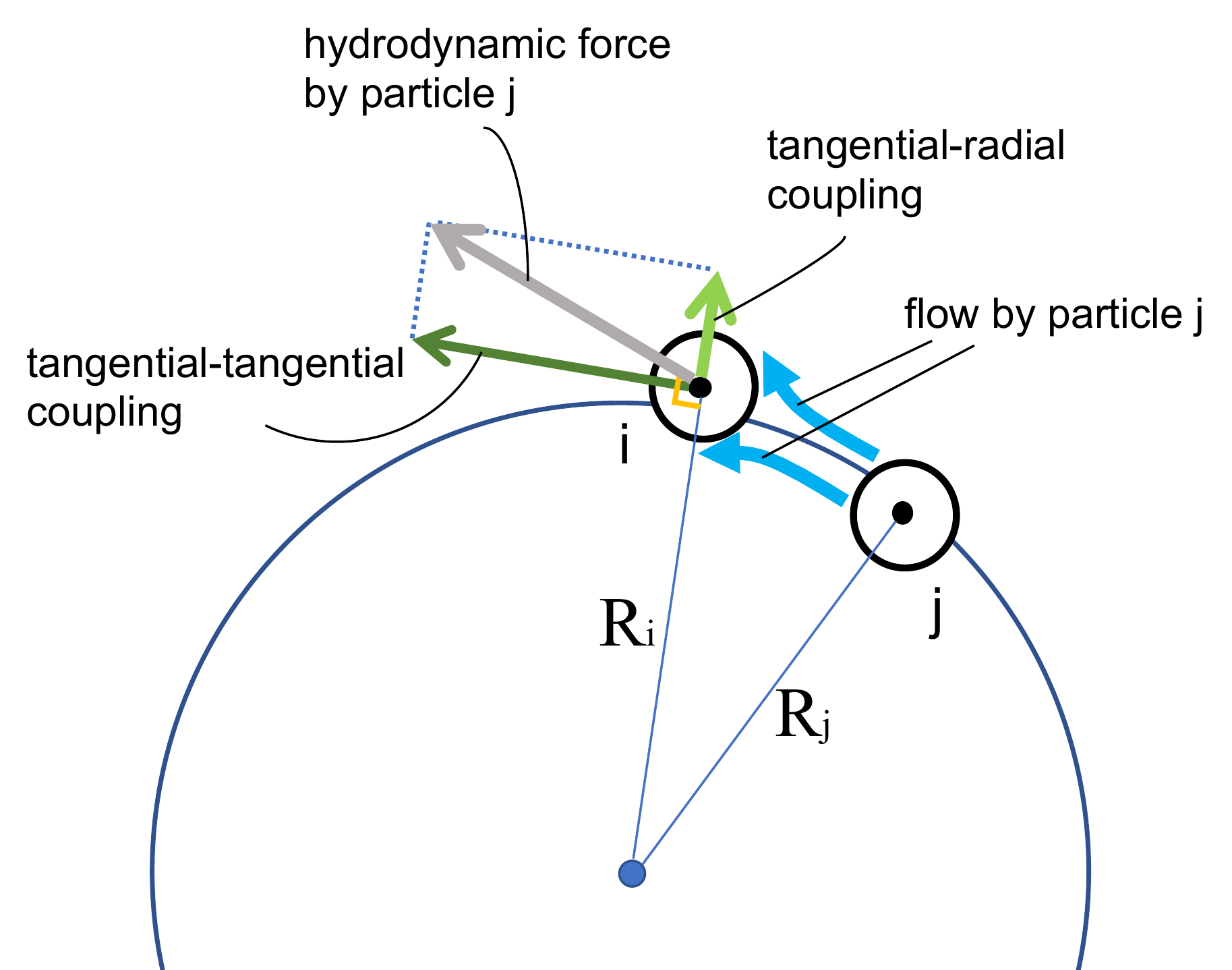}
\caption{\label{fig:coupling} 
Schematic picture of the hydrodynamic interaction. 
The $j$-th particle moving with the speed ${\bf v}_j \simeq R \omega_j {\bf e}^\phi_j$ 
generates the flow velocity ${\bf v}({\bf r}_i) \propto \hat{\textbf{G}}_{ij} \cdot {\bf e}^\phi_j$ at the position of the $i$-th particle.
The resultant hydrodynamic force is decomposed into the tangential component 
$\propto {\bf e}^\phi_i \cdot \hat{\textbf{G}}_{ij} \cdot {\bf e}^\phi_j$
and the radial component
$\propto {\bf e}^r_i \cdot \hat{\textbf{G}}_{ij} \cdot {\bf e}^\phi_j$.
The tangential-tangential coupling accelerates the front particle, 
while the tangential-radial coupling pushes it outward from the targetted circle (solid line)
and decreases the angular velocity. 
}
\end{figure}

\subsection{Phase Equation}

Now we derive the time evolution equation for the phase $\phi_i(t)$.
The equation of force balance is 
\begin{equation}
F_i^r \textbf{e}_i^r+F_i^\phi\textbf{e}_i^{\phi} + \sum_{j\sim i} {F}_{ji}^{rep} {\bf e}_i^\phi = 
\gamma \sum_{j \neq i} \left( \dot{\textbf{r}}_i - \hat{\bf{G}}_{ij} \cdot \dot{\textbf{r}}_j \right),
\label{balance}
\end{equation}
where $\sum_{j\sim i}$ means that the sum is taken over particles next to the $i$-th particle.
On the RHS, we have the hydrodynamic forces from (\ref{resistance}) and (\ref{vapprox}),
in which we substitute the particle velocity 
$
\dot{\bf r}_i = \dot{R}_i {\bf e}_i^r + R_i \dot{\phi}_i {\bf e}_i^\phi.
$
Taking the inner products of $\textbf{e}_i^r$, ${\bf e}_i^\phi$ and (\ref{balance}), 
we get
\begin{equation} 
- k_r \delta R_i
= 
\gamma \dot{R}_i - \gamma \sum_{j \neq i} \textbf{e}_i^r \cdot\hat{\textbf{G}}_{ij} \cdot \dot{\textbf{r}}_j 
\label{r}
\end{equation}
and 
\begin{equation}
\gamma R_i \dot{\phi}_i 
= 
F 
+ \textbf{e}_i^\phi \cdot \sum_{j\sim i }\textbf{F}_{ji}^{rep} 
+ \gamma  \sum_{j \neq i} \textbf{e}_i^\phi \cdot\hat{\textbf{G}}_{ij} \cdot \dot{\textbf{r}}_j,
\label{phi}
\end{equation}
respectively, where we also used (\ref{Fphi}) and (\ref{harmonic}). 
In the absence of hydrodynamic and repulsive interactions ($\alpha = u = 0$) 
and in the limit of strong trapping ($\kappa^{-1} = 0$),
Eq.(\ref{phi}) gives the intrinsic phase velocity
\begin{equation}
\omega = \frac{F}{\gamma R}. 
\label{omega}
\end{equation}
We also see from (\ref{r}) that the radial displacement is an $O(\kappa^{-1} \alpha)$ quantity. 
Then we can use $\dot{\bf r}_i \simeq R \dot{\phi}_i {\bf e}_i^\phi \simeq R \omega {\bf e}_i^\phi$
in the last term on the RHS of (\ref{r}), to obtain 
\begin{equation}
\frac{\delta R_i}{R} \approx 
\kappa^{-1}
\sum_{j\neq i} \textbf{e}_i^r \cdot \hat{\textbf{G} }_{ij}\cdot \textbf{e}_j^\phi.	
\label{deltaR_i}
\end{equation}
For Eq. (\ref{phi}), we use the same approximation on the RHS 
and divide by $\gamma \omega R_i = \gamma \omega R (1 + \delta R_i/R)$,  
in which (\ref{deltaR_i}) is substituted.
Neglecting terms of $O(\kappa^{-2} \alpha^2, \kappa^{-1} u, \alpha u)$, 
we arrive at the phase equation 
\begin{align}
\frac{\dot{\phi}_i }{\omega} = 1 &+ 
\sum_{j\neq i }\textbf{e}_i^\phi \cdot \hat{\bf G}_{ij} \cdot \textbf{e}_j^\phi 
- \kappa^{-1} 
\sum_{j\neq i }\textbf{e}_i^r \cdot \hat{\bf G}_{ij} \cdot \textbf{e}_j^\phi & \nonumber\\
&- 
\qty[ f^{rep}(\Delta_{i,(i+1){\rm mod} N }) - f^{rep}(\Delta_{(i-1){\rm mod} N,i}) ].&
\label{main}
\end{align}
Here, $\hat{\bf G}_{ij}$ is evaluated by setting $\delta R_i = \delta R_j = 0$
and becomes a function of the phases $\phi_i$ and $\phi_j$, 
and
\begin{equation}
f^{rep}(\Delta) = 6 u \left( \frac{|\Delta|}{2\alpha} - 1 \right)^{-13} 
\label{frep}
\end{equation}
is the dimensionless repulsive force. 
The hydrodynamic coupling is represented by the second and third terms on the RHS of (\ref{main}).
Their physical meanings are illustrated in Fig. \ref{fig:coupling} and are understood as follows. 
If two particles are moving in tandem, the particle in the rear pushes the front particle via the fluid 
and the front one pulls the rear. This effect is shown by the second term (tangential-tangential coupling)
and increases the phase velocity of the pair.  
However, because the particles are moving on a circle, the hydrodynamic force exerted by the rear 
causes a radial displacement of the front particle.
The increased orbital radius means decrease of the phase velocity as the driving force is constant.
This effect is shown by the third term (tangential-radial coupling).
The two terms are explicitly given by
\begin{equation}
\textbf{e}_i^r \cdot \hat{\bf G}_{ij} \cdot \textbf{e}_j^{\phi} =
- \frac{9 \alpha \sin\Delta_{ij} }
{16 \left|\sin\left(\frac{\Delta_{ij}}{2}\right)\right|}
+ \frac{\alpha^3 \sin\Delta_{ij}}
{32 \left|\sin\left(\frac{\Delta_{ij}}{2}\right)\right|^3}
, 
\end{equation}
and
\begin{equation}
\textbf{e}_i^{\phi} \cdot \hat{\bf G}_{ij} \cdot \textbf{e}_j^{\phi} 
= 
\frac{\alpha\qty(3 + 9\cos\Delta_{ij} )}
{16 \left|\sin\left(\frac{\Delta_{ij}}{2}\right)\right|}
- 
\frac
{\alpha^3  (3+\cos\Delta_{ij}) }
{32 \left|\sin\left(\frac{\Delta_{ij}}{2}\right)\right|^3}
.
\end{equation}

Note that the RHS of the phase equation (\ref{main}) are 
functions of the phase difference only,  and written in the form
\begin{equation}
\frac{\dot{\phi}_i }{\omega} = 1+ \sum_{k\neq i} \Omega(\Delta_{ik}).
\end{equation}
Therefore we can get a closed set of equations 
in terms of the phase difference, as
\begin{equation}
\frac{\dot{\Delta}_{ij} }{\omega} = \sum_{k \neq j} \Omega(\Delta_{jk}) - 
\sum_{k \neq i} \Omega(\Delta_{ik}),
\end{equation}
In particular, for $N=2$, the phase difference $\Delta_{12}$ obeys
\begin{eqnarray}
\frac{\dot{\Delta}_{12} }{\omega} &=& 
2\kappa^{-1} \textbf{e}_1^r \cdot \hat{\bf G}_{12} \cdot \textbf{e}_2^\phi + 2 f^{rep}(\Delta_{12})
\nonumber \\
&\equiv& - \frac{dV(\Delta_{12})}{d\Delta_{12}}
.
\label{eq:N=2}
\end{eqnarray}
The latter equality defines the effective potential $V(\Delta)$,
which is a useful tool to visualize the stability of the synchronized states~\cite{Uchida2011}.
It is calculated by integrating the RHS of (\ref{eq:N=2}), as 
\begin{eqnarray}
V(\Delta) =
\frac{9 \kappa^{-1}\alpha}{2} \sin \qty(\frac{\Delta}{2})
+
\frac{\kappa^{-1}\alpha^3}{4 \sin \qty(\frac{\Delta}{2} )}  
+ \frac{2u}{ \qty(\frac{\Delta}{2\alpha} - 1)^{12} }.
\nonumber\\
\end{eqnarray}

\subsection{Settings for Numerical Analysis}

\begin{figure}[h]
\includegraphics[clip, width=8.0cm]{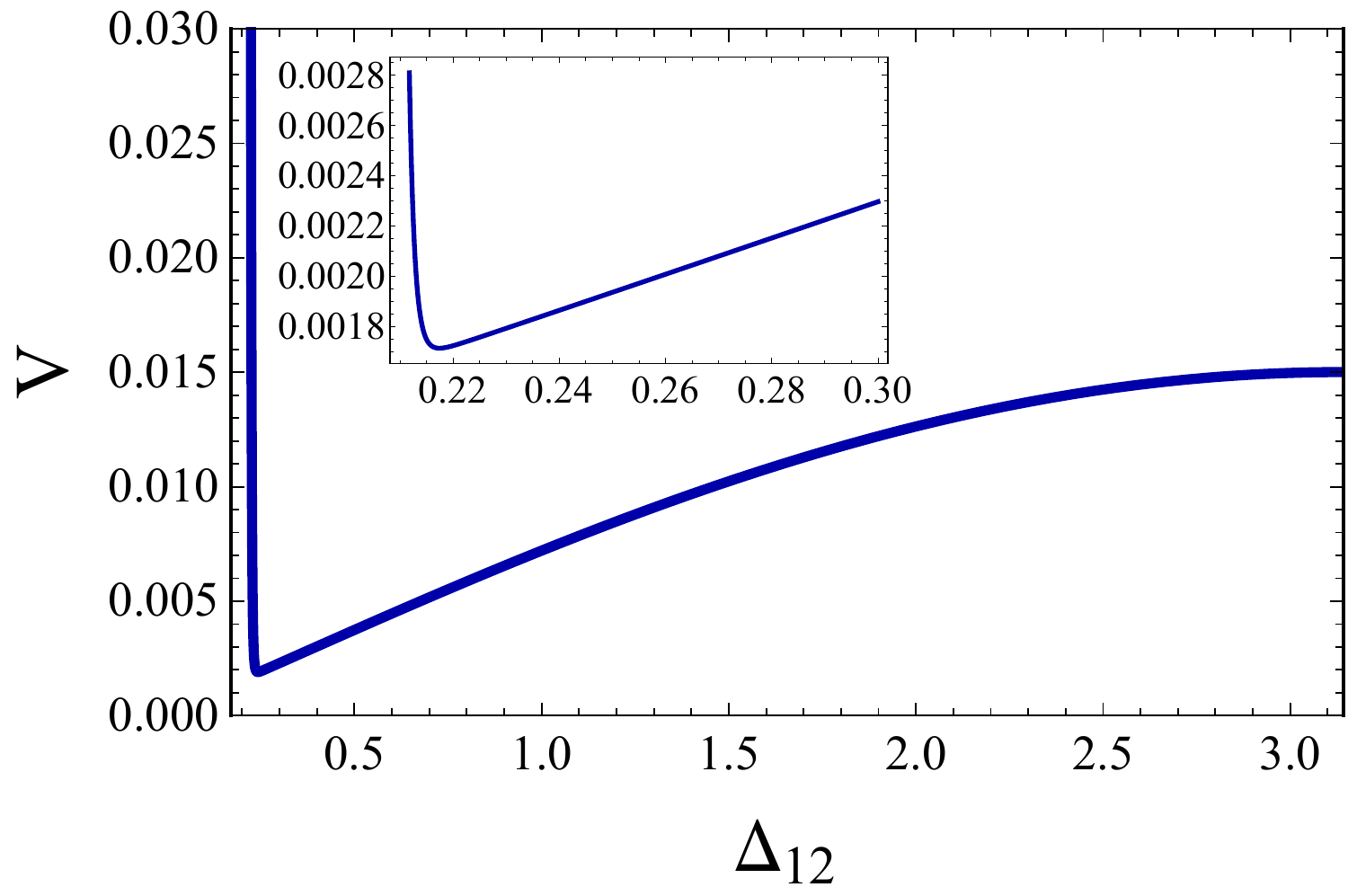}
\caption{\label{fig: N2_3} the effective potential for $\kappa=30, \alpha=0.1, u=10^{-12}$.  
It diverges at $\Delta_c = 2  \alpha= 0.2$.
(inset) Enlarged view around the minimum $\Delta_m\simeq 0.22$.}
\end{figure}

The model contains three small parameters $\kappa^{-1}= F/(k_r R)$, $\alpha = a/R$, and $u = U_0/(Fa)$.
In numerical analysis,
we used $\kappa = 30$, $\alpha = 0.1$, and $u = 10^{-12}$ unless stated otherwise.
The first two parameters are in the same order of magnitude as the experimental values
$\kappa = 30$ and $\alpha \simeq 0.24$ or $0.16$~\cite{Sokolov}.
The value of $u$ is chosen so that the repulsive force balances with the driving force ($U(\Delta) \simeq Fa$)
at the minimum of the effective potential $\Delta = \Delta_m \simeq 0.22$.
The effective potential is plotted in Fig. \ref{fig: N2_3} with these parameters.
The repulsive potential for $\Delta < \Delta_m$ is very steep and 
prevent the particles to contact each other, while the hydrodynamic coupling 
causes a long-range attraction.

In the following, we take $\omega^{-1}$ as the unit of time so that 
the intrinsic phase velocity $\omega$ is unity.
We used {\it Mathematica} to numerically integrate the phase equations.

\section{\label{sec:level3}Numerical Results}

\subsection{The case $N = 2$}

\begin{figure}[h]
\includegraphics[clip, width=9.0cm]{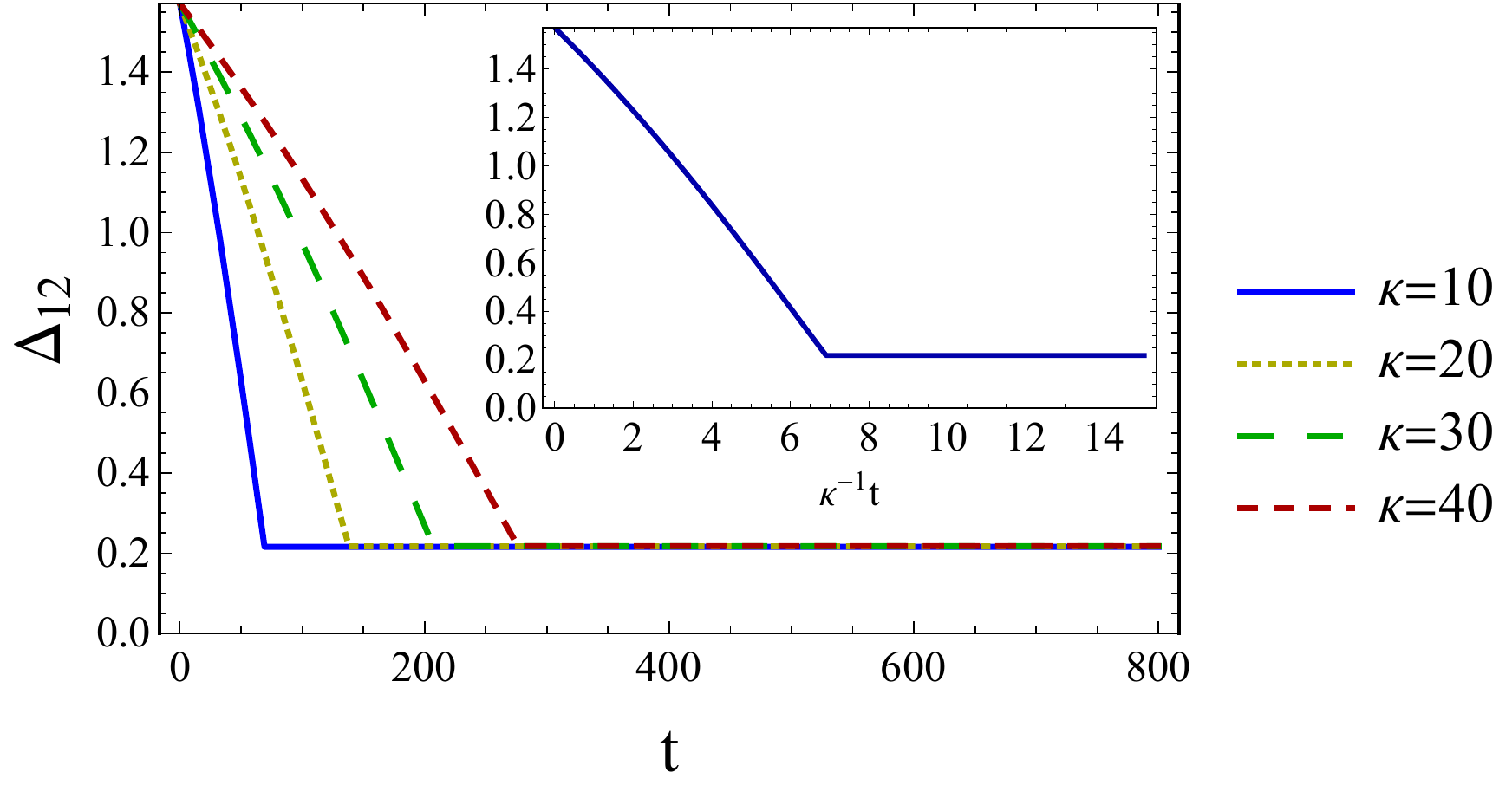}
\caption{\label{fig: N2_10} 
time evolution of the phase difference $\Delta_{12}$ 
for $N=2$ and for different values of $\kappa$. 
(inset) Master plot with the scaled time $\kappa^{-1} t$) }
\end{figure}

For an $N=2$ system ,
we show the time evolution of $\Delta_{12}$ in Fig. \ref{fig: N2_10} 
with $\kappa$ varied and the initial condition $\Delta_{12} = \pi / 2$.
The phase difference monotonically decrease to 
the minimum of the effective potential, where a doublet is formed.
As the radial confinement by the optical vortex becomes stronger,
it takes more time to form a doublet. 
The master plot with the scaled time $\kappa^{-1} t$ in the inset of  Fig. \ref{fig: N2_10}
shows that the timescale of  doublet formation is proportional to $\kappa$.
This is understood from Eq.(\ref{eq:N=2}) and the fact that 
the repulsive force is  negligibly small except near contact.

\begin{figure}[h]
\includegraphics[clip, width=8.0cm]{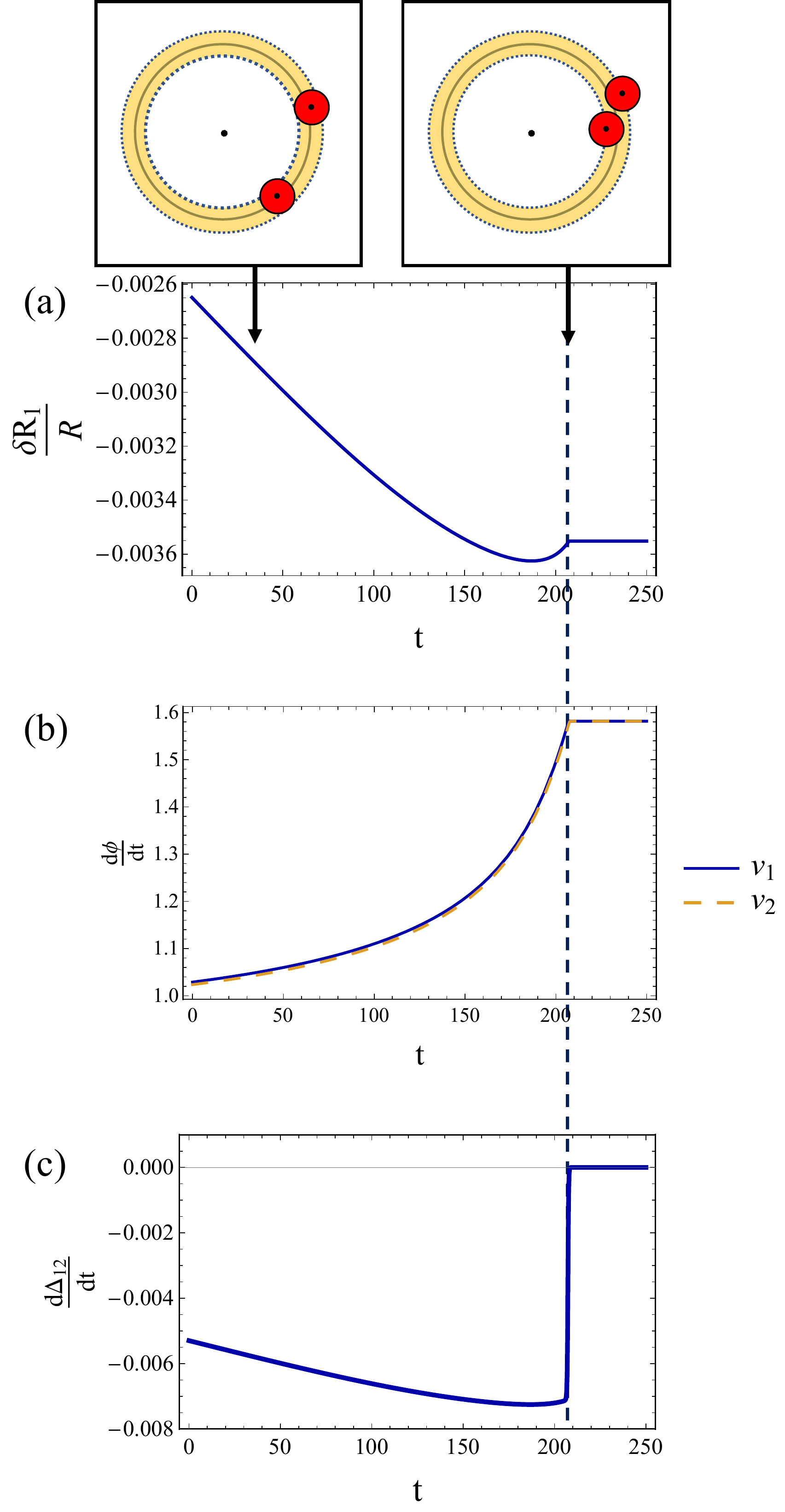}
\caption{\label{fig: N2_12} 
Time evolution of 
(a) the radial displacements, 
(b) phase velocities, 
and 
(c) the difference of the phase velocities for an $N=2$ system.
The two particles form doublet and synchronized at the time indicated by the dotted line.
Particle configurations at the time indicated by the arrows
are shown schematically.
}
\end{figure}

The time evolution of the radial displacement and phase velocity 
are shown in Fig. \ref{fig: N2_12}.
As we see from the figure,  the radial displacement of the front particle (particle 2)
increases as the two particles approach each other
and shows a small drop before it converges to a constant.
The displacement of the rear particle (particle 1)  is $\delta R_1 = - \delta R_2$
as seen from (\ref{deltaR_i}).
As the radial displacements are much smaller than the particle radius 
($|\delta R_i/R| \simeq 0.004 \ll \alpha=0.1$),
we can justify the assumption that the repulsive force 
acts only in the tangential direction.

The phase velocities of the two particles  monotonically increase
and converge to a constant 
\begin{equation}
v_d \simeq 1.58, 
\end{equation}
which is the velocity of the doublet.
Because the velocity of a singlet in the $N = 1$ case is set to unity ($v_s= 1$)  
by assumption,  the ratio between the doublet and singlet velocities is 
$v_d/v_s \simeq 1.58$. 

A doublet is stable because the rear particle pushes the front particle out 
from the target trajectory by hydrodynamic interaction, 
which makes the driving torque smaller for the front particle~\cite{Sokolov}.

\subsection{The case $N=3$}

\begin{figure}[hh]
\includegraphics[clip, width=8.0cm]{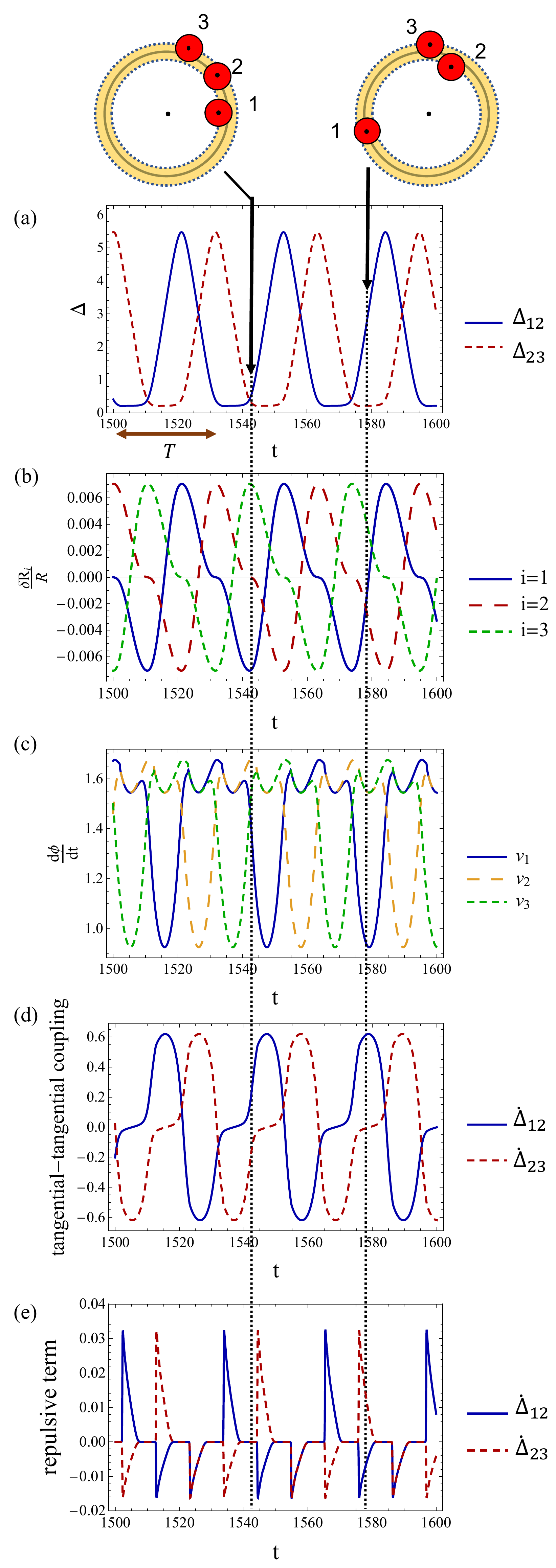}
\caption{\label{fig: N3_2} 
Time evolution of an $N=3$ system:
(a) Phase differences.
(b) Radial displacements.
(c) Phase velocities.
Contributions to $\dot\Delta_{12}$ and $\dot\Delta_{23}$ 
from 
(d) the tangential-tangential coupling term 
and 
(e) the repulsive term
in the phase equation (\ref{main})
are shown by the solid and dashed lines, respectively.
Particle configurations at the time indicated by the arrows
are shown schematically.
}
\end{figure}

The solution of the phase equation (\ref{main}) for an $N=3$ system 
becomes a periodic function of time, as shown in Fig. \ref{fig: N3_2}.
At the time indicated by the left arrow, 
the particles form a triplet (labeled 1-2-3) as we have $\Delta_{12} = \Delta_{23} \simeq 2 \Delta_m$.
At the time indicated by the right arrow,
the particles are separated into a doublet and a singlet 
as we have $\Delta_{12} \gg \Delta_{23} \simeq \Delta_m$.
The doublet is faster than a singlet and 
goes one more cycle than the singlet to form a triplet again.  
Three collisions take place in a period $T$
to shuffle the triplets, 1-2-3, 3-1-2, and 2-3-1.
The period is found to be 
\begin{equation}
T_{N=3} \simeq 30.6.
\label{T}
\end{equation}

Time evolution of the reduced radial displacements $\delta R_i/R$
are shown in Fig. \ref{fig: N3_2} (b).
In a triplet, the front (rear) particle has positive (negative) radial displacement, respectively,
while the center particle is located on targeted trajectory,
as shown schematically in the figure. 

The phase velocity of each particle is plotted in Fig. \ref{fig: N3_2} (c).
It has three peaks and one deep valley in one cycle.
Let us focus on the particle 1 in the plot.
The first peak is obtained when the particle 
is collided at its rear end (by the particle 3) and form a doublet.
The second and the highest peak is reached when 
the doublet collides with a singlet and the particle 1 becomes
the center of the  triplet 3-1-2.
The triplet is split into the doublet 1-2 and the singlet 3
and the phase velocities decrease.
The third and the lowest peak is attained
when the doublet 1-2 approaches the singlet 3.
The minimal phase velocity is marked by a singlet 
when it is separated from the doublet and 
is on the opposite side of the circle.

The doublet phase velocity $v_{d}$ is calculated 
by averaging over the time window 
during which the particle 1 forms a doublet with the particle 3, as 
\begin{equation}
v_{d} \simeq 1.57.
\label{v}
\end{equation}
The period (\ref{T}) is estimated as the time 
for  a doublet to go one more cycle than a singlet, 
multiplied by three to account for the three shuffles, 
as $T_{N=3} \simeq 2\pi / (v_d - v_s ) \times 3\simeq 33.1$,
which is only 10\% larger than the value (\ref{T}).

Now let us consider the reason why a triplet is unstable and split~\cite{Reichert2004, Sassa}. 
In Fig.\ref{fig: N3_2} (d)(e), we show the contributions to $\dot\Delta_{ij}$ from 
the tangential-tangential coupling (the second term on the RHS of Eq.(\ref{main}))
and the repulsive force, respectively.
Compared to them, the contribution of the tangential-radial coupling 
(the third term on the RHS of Eq.(\ref{main})) is negligibly small  (not shown).
Thus the splitting of a triplet is caused by the tangential-tangential coupling,
which is the dominant contribution to $\dot\Delta_{ij}$.
As seen from Fig. \ref{fig: N3_2}(d),  
the phase differences satisfy $\dot{\Delta}_{12} \sim -\dot{\Delta}_{23} > 0$
when the triplet $1-2-3$ is formed.
This is because the center particle 2 is hydrodynamically screened and 
receives a drag force weaker than those on the other particles~\cite{Reichert2004}.
Thus the triplet is split into a doublet and a singlet.
On the other hand, the repulsive force shows a sharp peak
when the particles forming a doublet get closest.
However, its contribution to $\dot\Delta_{ij}$ is 
an order of magnitude smaller than the tangential-tangential coupling,
as we see from Fig. \ref{fig: N3_2}(e).

\begin{figure}[h]
\includegraphics[clip, width=7.5cm]{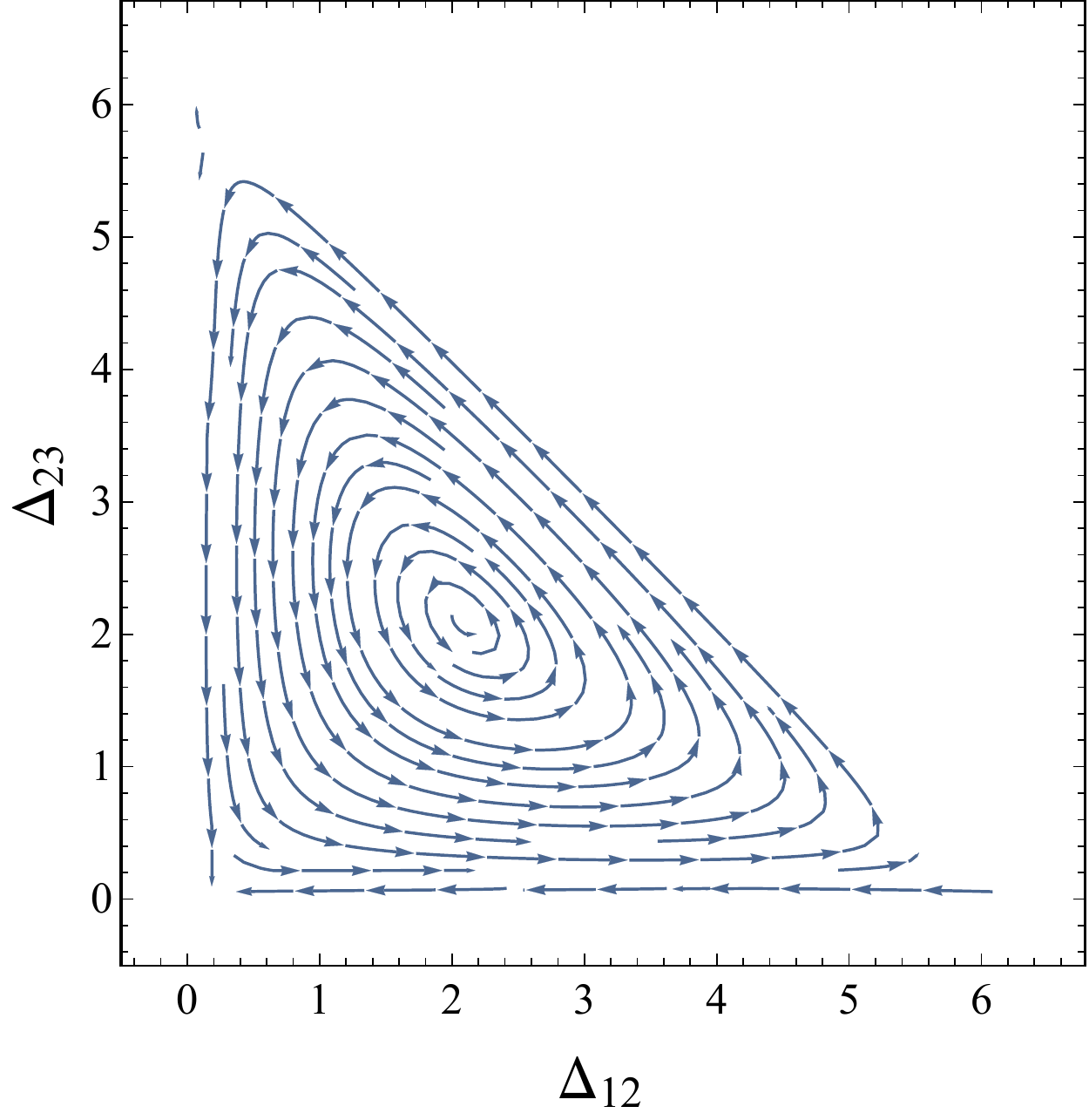}
\caption{\label{fig: N3_9} 
Phase portrait in the $\Delta_{12}-\Delta_{23}$ plane for an $N=3$ system.
}
\end{figure}

The phase portrait of the dynamics is given in terms of 
the trajectory in the  $\Delta_{12}$-$\Delta_{23}$ plane, 
which is shown in Fig.\ref{fig: N3_9}.
In the long-time limit, the trajectory converges to a triangle-shaped limit cycle.
In other words, the system exhibits the same rhythmic motion 
regardless of the initial conditions.

\subsection{The case $N=4$}

\begin{figure}[h]
\includegraphics[clip, width=9.0cm]{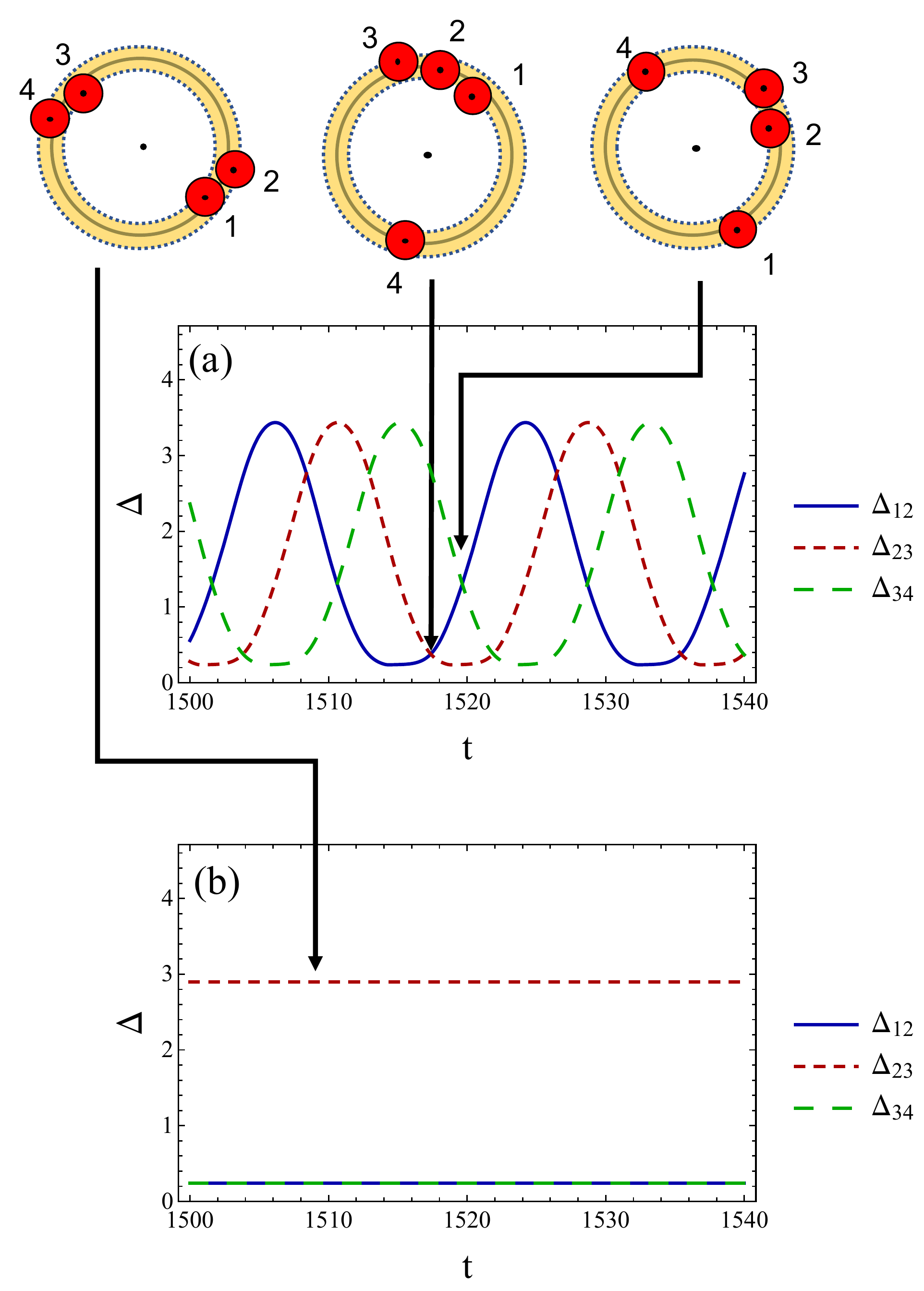}
\caption{\label{fig: N4_2} 
Time evolution of the phase differences in an $N=4$ system
for (a) the oscillating state and (b) the stable state. 
Particle configurations at the time indicated by the arrows
are shown schematically.
}
\end{figure}

The time evolution of an $N=4$ system is shown in Fig. \ref{fig: N4_2}.
Depending on the initial configurations,
we obtained either 
(i) a periodically oscillating state consisting of 
a doublet and two singlets that are shuffled with each other, 
or 
(ii) a stable synchronized state consisting of two doublets.

The period for the oscillating state is found to be
\begin{equation}
T_{N=4} \simeq 17.1,
\label{T4}
\end{equation}
which is much smaller than the period for $N=3$.
This is because two singlets are positioned 
in the opposite side of the circle 
when a doublet moves between the two. 
Therefore, the distance traveled by a doublet in an $N=4$ system 
between two collisions is roughly half of the one for $N=3$.
Taking into account the four shuffles per period,
and using the doublet velocity $v_d=1.63$,
we get the estimate
$T_{N=4} \simeq \pi/(v_d- v_s) \times 4 \simeq 19.9$.

For the stable synchronized states, 
the phase difference between two doublets is $\Delta_{23} = \Delta_{41} \simeq 0.92\pi$,
which means that the doublets are placed in the opposite side of the circle.

\begin{figure}[h]
\includegraphics[clip, width=9.5cm]{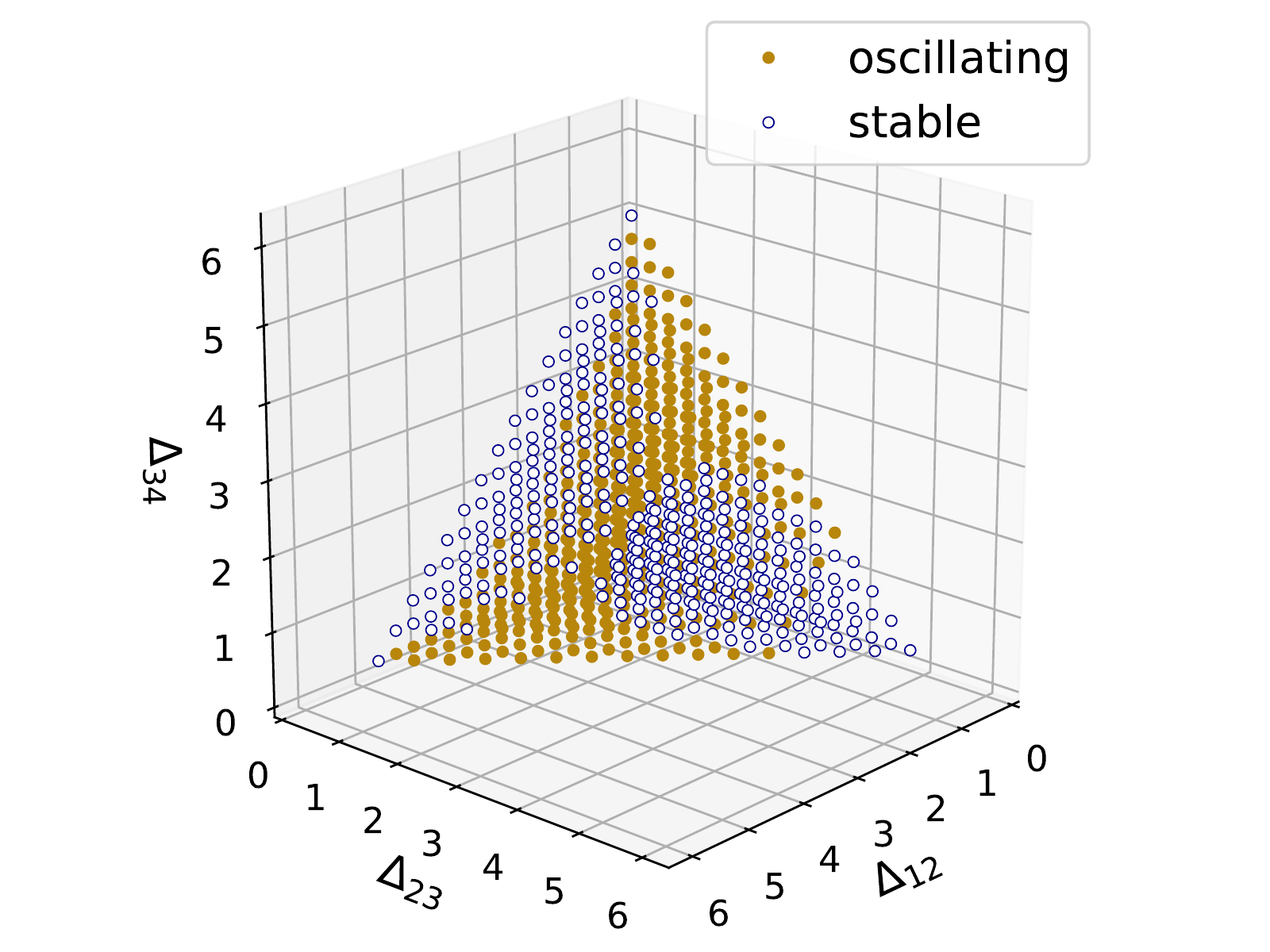}
\caption{\label{fig: N4_10} 
State diagram for $N=4$, showing dependence on the initial conditions
in terms of the phase differences.
The stable state is obtained from relatively symmetric initial configurations.
The oscillating state is obtained if two particles are close in the initial state.
}
\end{figure}

In Fig. \ref{fig: N4_10}, we show the state diagram in terms of the initial conditions.
The initial conditions are varied by setting $\phi_i$'s $(i=1,2,3,4)$ as integer multiples of $\epsilon=0.1 \pi$.
The state diagram is drawn in the $(\Delta_{12}, \Delta_{23}, \Delta_{34})$ space
with $\Delta_{ij}$'s ranging in $[\epsilon, 16 \epsilon]$.
It shows that 
the system reaches the stable state if two of the particles are close to each other 
in the initial state.
For example, the corner points of the state diagram,
$(\Delta_{12},\Delta_{23}, \Delta_{34}) = (16\epsilon,\epsilon,\epsilon), (\epsilon,16\epsilon,\epsilon),(\epsilon,\epsilon,16\epsilon)$,
give the stable state.
The oscillating state is obtained if the particles are equally spaced or 
form a triplet and a singlet in the initial state.
The number fraction of the initial configurations that give rise to the oscillating state was 54\%. 
Therefore, it is slightly more probable to get the oscillating state
if we randomly choose the initial configuration.
%

\subsection{The case $N=5$}

\begin{figure}[h]
\includegraphics[clip, width=9.0cm]{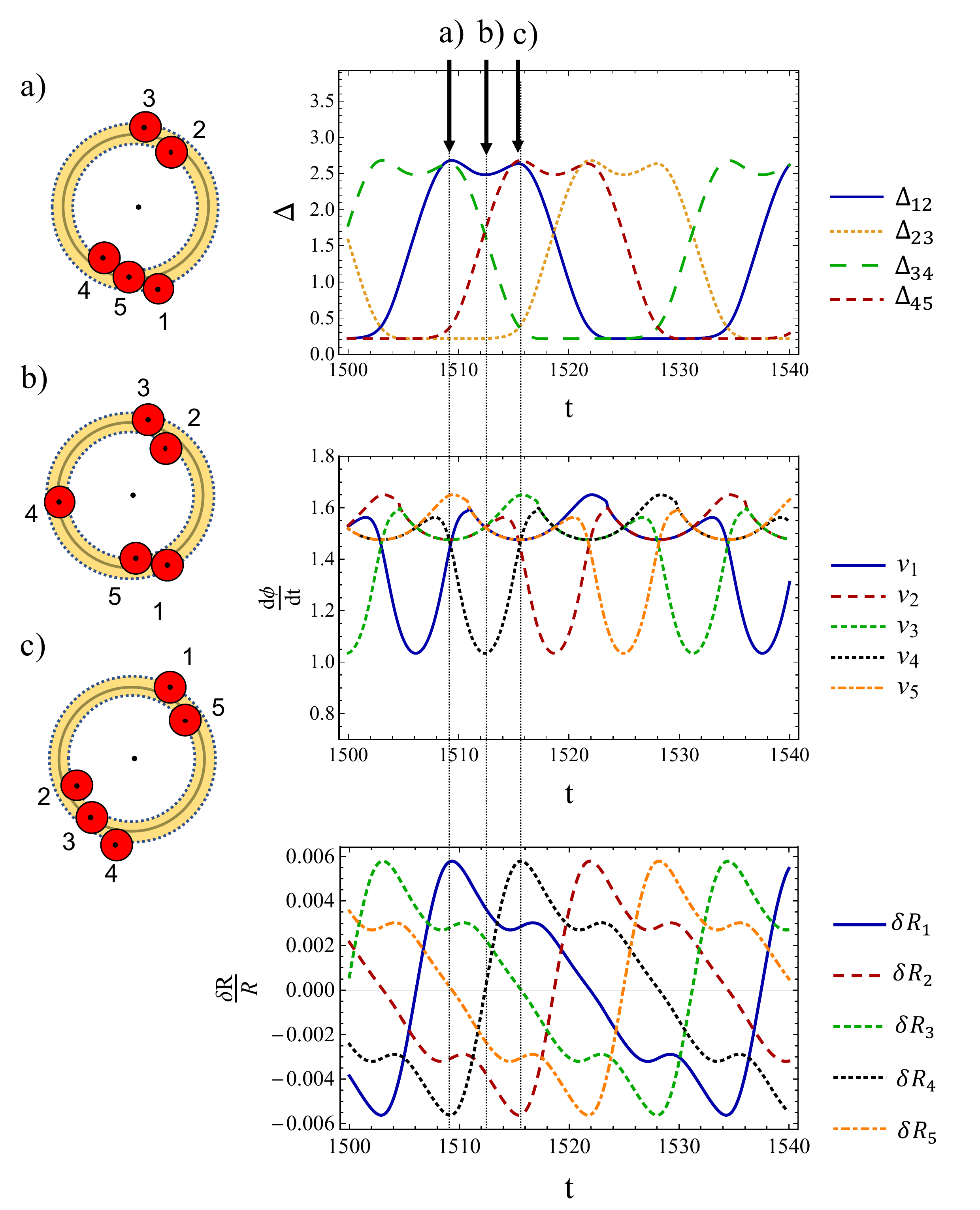}
\caption{\label{fig: N5_1}
Time evolution of an $N=5$ system.
(Top) Phase differences. 
(Middle) Radial displacements. 
(Bottom) Phase velocities. 
Particle configurations at the time indicated by the arrows a-c
are shown schematically in a)-c), respectively.
}
\end{figure}

In Fig. \ref{fig: N5_1}, we show the time evolution of 
the phase difference, radial displacements and phase velocities
for an $N=5$ system.
As in the case $N=3$, the system converges to 
a periodic oscillating state independent of initial conditions. 
The value of the period is almost the same as in the case $N=3$:
\begin{equation}
T_{N=5} \simeq 31.4.
\end{equation}
The phase difference exhibit double peaks in a period, 
which is a feature not found in the cases $N=3,4$.
This is understood as follows.
At the first peak marked by the arrow a) in the figure,
the triplet 4-5-1 and the doublet 2-3 are formed.
Then the triplet moves faster than the doublet
as seen in Fig. \ref{fig: N5_1} (middle),
because the center particle is hydrodynamically screened~\cite{Reichert2004,Sassa}.
The triplet is decomposed  
into  a doublet and a singlet at the time indicated by the arrow b).
The new doublet (5-1) is slower than the other doublet (2-3) 
because it has a larger radial displacement, 
as seen from Fig. \ref{fig: N5_1} (bottom).
This causes the slight decrease of the phase difference $\Delta_{12}$
and the dip between the two peaks.
Then the triplet 2-3-4 is formed at the time indicated by the arrow c),
and $\Delta_{12}$ increases again to second peak.

\subsection{The case $N=6$}

\begin{figure}[b]
\includegraphics[clip, width=9.0cm]{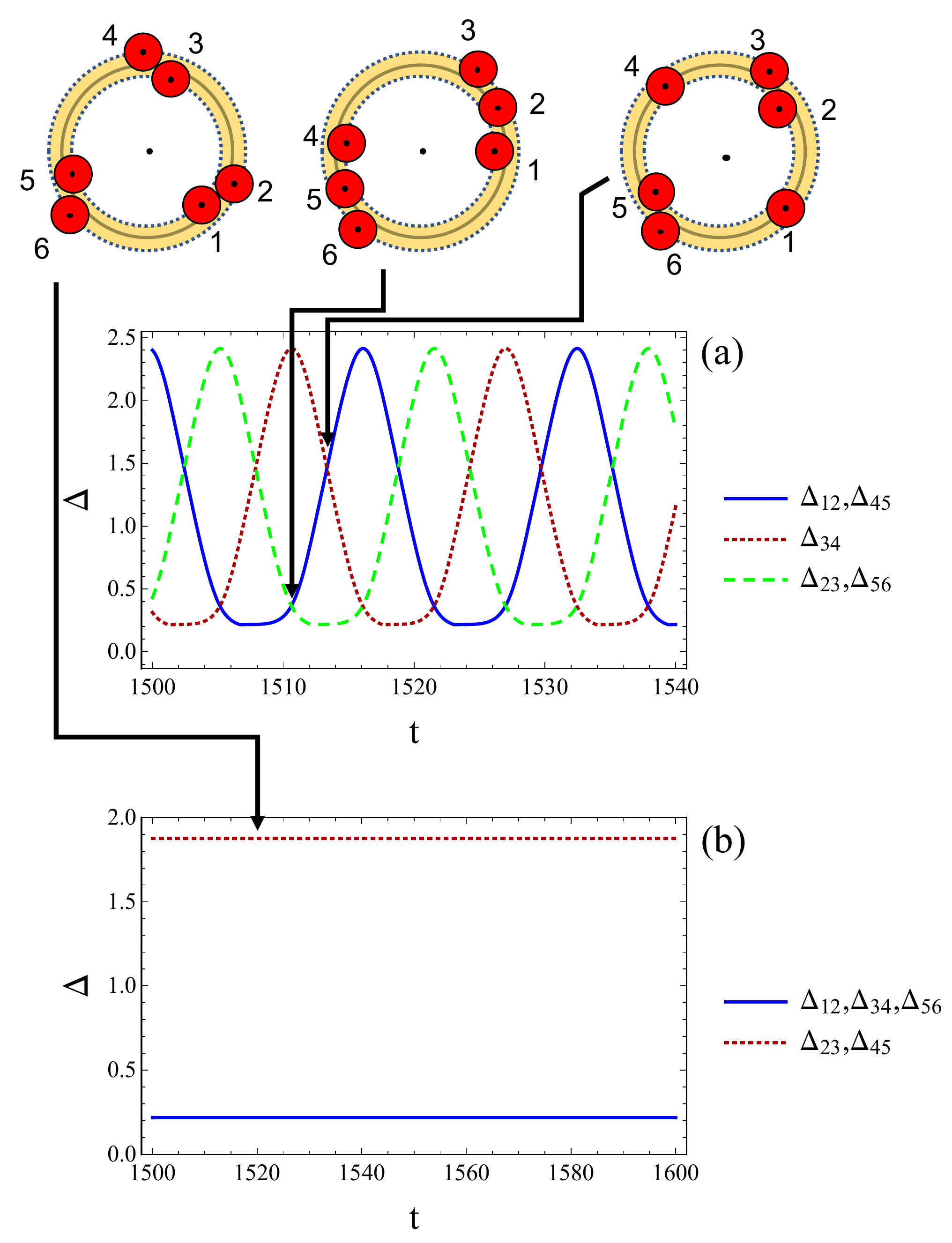}
\caption{\label{fig: N6_2} 
Time evolution of the phase differences in an $N=6$ system for 
(a) the oscillating state and (b) the stable state.  
Particle configurations at the time indicated by the arrows 
are shown schematically.
}
\end{figure}

Time evolution of the phase differences for an $N=6$ system is shown 
in Fig.\ref{fig: N6_2}.
As in the case $N=4$, 
the system converges to either an oscillating or a stable synchronized states,
depending on the initial conditions.
The oscillating state consists of two doublets and two singlets.
The period is almost the same as in the case $N=4$:
\begin{equation}
T_{N=6} \simeq 16.4.
\end{equation}
The stable state consists of three doublets that are equally spaced.


\begin{table*}
\caption{\label{vel.}
The maximum, minimum and time-averaged phase velocities $v_{\rm max}, v_{\rm min}, v_{\rm ave}$ 
and  period of oscillation $T$.
For $N=4$ and $6$, we show the values for both stable and oscillating (osc.) modes. 
}
\begin{ruledtabular}
\begin{tabular}{ccccccccc}
& $N=2$ & $N=3$ & $N=4$(osc.) & $N=4$(stable)  & $N=5$ & $N=6$(osc.) & $N=6$(stable) \\
\hline
$v_{\rm max}$ & 1.58 & 1.67 & 1.68 & 1.46 & 1.65 & 1.64 & 1.54  \\
$v_{\rm min}$ & 1.58 & 0.93 & 0.89 & 1.46 & 1.03 & 1.07 & 1.54 \\
$v_{\rm ave}$ & 1.58 & 1.42 & 1.36 & 1.46 & 1.46 & 1.45 & 1.54 \\
$T$ & - & 31.6 & 18.1 & - & 31.4 & 16.4 & - \\
\end{tabular}
\end{ruledtabular}
\end{table*}

\section{\label{sec:level4}Discussion}

Now let us discuss the results in comparison with
the previous studies and consider their physical meanings.
A quantity of interest is the ratio between the phase velocities of 
a doublet and a singlet, $v_d/v_s$.
It is always larger than unity because the hydrodynamic resistance is 
reduced as two particles get close. 
For an $N=2$ system, we can estimate the ratio analytically by 
considering two spheres that are in contact with each other 
and moving in tandem along a straight line.
The Rotne-Prager-Yamakawa approximation (\ref{two_body_approx.}) 
gives $v_d/v_s = 8/5 = 1.6$, 
which is close to our numerical result $v_d/v_s \simeq 1.58$.
The small discrepancy is resulting from the non-vanishing 
radial displacement and tangential separation in the numerical analysis.
The Oseen tensor, which is obtained by neglecting 
the correction terms in the Rotne-Prager-Yamakawa tensor due to finite particle radius,
gives the much smaller value $v_d/v_s = 4/3 \simeq 1.33$.
A more exact value  is obtained from the lubrication theory~\cite{Kim} 
as $v_d/v_s \simeq 1.578$. 
This is fairly close to the value from the Rotne-Prager-Yamakawa approximation
as well as to our numerical result.
The finite temperature simulations by Sokolov et al.~\cite{Sokolov} 
gave $v_d/v_s = 1.57\pm 0.05$, which is also close to our result. 
Experimentally, the ratio $v_d/v_s$ is reported as $1.10 \pm 0.02$ in Ref.\cite{Sokolov} 
and $1.32$ by Sassa {\it et al.}~\cite{Sassa}.
The reason why the experimental values are smaller than the theoretical ones
is not clear at this stage. However, it is reported in Ref.~\cite{Sassa} 
that fluctuations in the singlet velocity is as large as 27\%,
which is attributed to nonuniformity of the optical driving force along the circular path.
In an earlier experiment~\cite{Roichman}, the fluctuations of the optical intensity 
along the trajectory is also about 20\% from the mean.
If fluctuations in the radial displacement is large, 
they would cause a larger viscous drag force on a doublet
because the hydrodynamic screening by the front particle is weakened.

For $N \ge 3$, we confirmed the existence of oscillating modes with mutually exchanging 
doublets and singlets.
The limit cycle is robust and obtained from any initial conditions.
This is in tune with the experimental finding that 
a triplet collapses in some occasions but is soon recovered~\cite{Sassa}.

For $N=3$, the oscillating state is  described by a limit cycle in the phase portrait.
Similar phase portraits are obtained in the experiment~\cite{Kimura2017} 
and in the direct numerical simulation~\cite{Okubo2015}.
However, the trajectories in the experiment show noisy fluctuations,
which are probably due to the fluctuations of the driving force.
The trajectories obtained from the simulation also show oscillations,  
which should be due to periodic variation of the driving force assumed in the model.

For $N=4$ and $N=6$, 
both oscillating and stable states were observed in the experiment~\cite{Sassa}
and in the simulation~\cite{Sokolov},
in agreement with our results. According to our analysis, 
one of the two states is finally chosen depending on the initial conditions.
However, in the experiment, transition between the two states in a time course 
was observed, and finally only the stable states remained.  
This transition may be also attributed to fluctuations of the optical driving force~\cite{Sassa}.
There are not only quenched heterogeneities but also temporal fluctuations 
of the optical force.
In the simulation with thermal noise~\cite{Sokolov}, 
the oscillating mode for $N=4$ disappeared at temperature higher than the experimental temperature.
Temporal fluctuations of the optical force may be described by 
high effective temperature and explain the transition to the stable mode.

For $N=5$, we found an oscillating state with two doublets and a singlet, 
in accordance with experiments~\cite{Sassa}.
In the experiment,  another oscillating mode with one stable doublet and three singlets  
is also observed sometimes and in a short time period. 
An oscillating mode with one doublet is also observed for $N=6$.
These modes containing one doublet  is not reproduced in our analysis, 
and is possibly due to fluctuations of the optical driving force.

In Table~\ref{vel.}, we show
the maximal, minimal and time-averaged phase velocities 
as well as the periods of the oscillating modes,
all obtained in the dynamical steady states.
The phase velocities for the stable modes ($N=2, 4$ (stable),  $6$ (stable) in Table~\ref{vel.})
are those of a doublet.
For oscillating modes for $N=3, 4$ (osc.), $5, 6$ (osc.),
the maximal phase velocity is realized by the front particle of a doublet 
when the doublet collides with a singlet and a triplet is formed.
The minimal phase velocity is realized by a singlet
when it is separated from other particles. 
For example, for $N=3$, the singlet is slowest 
when the doublet is in the opposite side of the circle,
because the doublet causes a hydrodynamic drag force that 
is anti-parallel to the moving direction of the singlet. 
Therefore the deviation of the minimal velocity 
from the reference value (unity) is in the order of $\alpha$.

The mean phase velocity can be compared with 
those reported in the preceding studies~\cite{Sokolov, Sassa}. 
The mean phase velocity for an $N=2n$ system  in the stable mode 
is larger than that for $N=2n - 1$  ($n = 1, 2, 3$) .
This even-odd effect is reported in the experiments~\cite{Sokolov,Sassa}
and is interpreted by the number of doublets: for $N=2n$, 
The mean phase velocity for $N=2$ is larger than that for $N=4$ (stable).
This also agrees with the experimental results~\cite{Sokolov,Sassa}.
We interpret it as effect of the hydrodynamic drag force between the doublet 
and the two singlets that are moving in the opposite side of the circle,
just as in the above explanation of the minimal singlet velocity. 
On the other hand, 
the mean phase velocity for an $N=4$ system in the oscillating mode 
is smaller than that for $N=3$,
which is consistent with the previous simulation result~\cite{Sokolov}.
As a function of $N$, the mean phase velocity took minimum value 
at $N = 3$ in the experiments~\cite{Sokolov, Sassa}.
However, we found that the minimum is achieved by an $N=4$ system in the oscillating mode, 
which  is not stable  in the experiments as mentioned above.

In conclusion, we analyzed theoretically the collective dynamics of 
colloidal particles driven by an optical vortex.
We derived the phase equation for $N$ particles and 
quantitatively investigated the mechanisms behind 
the rich dynamical behaviors, such as:
(i) the radial degree of freedom for doublet formation,
(ii) hydrodynamic screening for triplet splitting,
and 
(iii) symmetry of particle configurations determining the basin of attractors.
The results are in good agreement with the experimental findings, 
including the even-odd effects in the phase velocity.
A remaining problem is an understanding of the transitions 
between the oscillating and stable modes observed in the experiments,
which might be related to spatio-temporal fluctuations of the driving force. 
In future experiments, 
a more precise control of the fluctuations of the optical force 
will be useful in unveiling the mechanism of the transitions,
and also in quantitative assessment of the phase velocities.
Theoretically, both spatial and temporal fluctuations of the driving force will be 
incorporated using the existing framework~\cite{Uchida2011,Maestro2018}.
The effects of confining walls \cite{Saito2018} and polydispersity of 
the particle size \cite{Okubo2015} also remain to be analyzed in future work.


\begin{thebibliography}{31}%
\makeatletter
\providecommand \@ifxundefined [1]{%
 \@ifx{#1\undefined}
}%
\providecommand \@ifnum [1]{%
 \ifnum #1\expandafter \@firstoftwo
 \else \expandafter \@secondoftwo
 \fi
}%
\providecommand \@ifx [1]{%
 \ifx #1\expandafter \@firstoftwo
 \else \expandafter \@secondoftwo
 \fi
}%
\providecommand \natexlab [1]{#1}%
\providecommand \enquote  [1]{``#1''}%
\providecommand \bibnamefont  [1]{#1}%
\providecommand \bibfnamefont [1]{#1}%
\providecommand \citenamefont [1]{#1}%
\providecommand \href@noop [0]{\@secondoftwo}%
\providecommand \href [0]{\begingroup \@sanitize@url \@href}%
\providecommand \@href[1]{\@@startlink{#1}\@@href}%
\providecommand \@@href[1]{\endgroup#1\@@endlink}%
\providecommand \@sanitize@url [0]{\catcode `\\12\catcode `\$12\catcode
  `\&12\catcode `\#12\catcode `\^12\catcode `\_12\catcode `\%12\relax}%
\providecommand \@@startlink[1]{}%
\providecommand \@@endlink[0]{}%
\providecommand \url  [0]{\begingroup\@sanitize@url \@url }%
\providecommand \@url [1]{\endgroup\@href {#1}{\urlprefix }}%
\providecommand \urlprefix  [0]{URL }%
\providecommand \Eprint [0]{\href }%
\providecommand \doibase [0]{https://doi.org/}%
\providecommand \selectlanguage [0]{\@gobble}%
\providecommand \bibinfo  [0]{\@secondoftwo}%
\providecommand \bibfield  [0]{\@secondoftwo}%
\providecommand \translation [1]{[#1]}%
\providecommand \BibitemOpen [0]{}%
\providecommand \bibitemStop [0]{}%
\providecommand \bibitemNoStop [0]{.\EOS\space}%
\providecommand \EOS [0]{\spacefactor3000\relax}%
\providecommand \BibitemShut  [1]{\csname bibitem#1\endcsname}%
\let\auto@bib@innerbib\@empty
\bibitem [{\citenamefont {Bray}(2001)}]{bray2001cell}%
  \BibitemOpen
  \bibfield  {author} {\bibinfo {author} {\bibfnamefont {D.}~\bibnamefont
  {Bray}},\ }\href {https://books.google.co.jp/books?id=yd61229NHUgC} {\emph
  {\bibinfo {title} {Cell Movements: From Molecules to Motility}}}\ (\bibinfo
  {publisher} {Garland Pub.},\ \bibinfo {year} {2001})\BibitemShut {NoStop}%
\bibitem [{\citenamefont {Maestro}\ \emph {et~al.}(2018)\citenamefont
  {Maestro}, \citenamefont {Bruot}, \citenamefont {Kotar}, \citenamefont
  {Uchida}, \citenamefont {Golestanian},\ and\ \citenamefont
  {Cicuta}}]{Maestro2018}%
  \BibitemOpen
  \bibfield  {author} {\bibinfo {author} {\bibfnamefont {A.}~\bibnamefont
  {Maestro}}, \bibinfo {author} {\bibfnamefont {N.}~\bibnamefont {Bruot}},
  \bibinfo {author} {\bibfnamefont {J.}~\bibnamefont {Kotar}}, \bibinfo
  {author} {\bibfnamefont {N.}~\bibnamefont {Uchida}}, \bibinfo {author}
  {\bibfnamefont {R.}~\bibnamefont {Golestanian}},\ and\ \bibinfo {author}
  {\bibfnamefont {P.}~\bibnamefont {Cicuta}},\ }\href
  {https://doi.org/10.1038/s42005-018-0031-6} {\bibfield  {journal} {\bibinfo
  {journal} {Communications Physics}\ }\textbf {\bibinfo {volume} {1}},\
  \bibinfo {pages} {28} (\bibinfo {year} {2018})}\BibitemShut {NoStop}%
\bibitem [{\citenamefont {Bruot}\ and\ \citenamefont
  {Cicuta}(2016)}]{Bruot-Cicuta}%
  \BibitemOpen
  \bibfield  {author} {\bibinfo {author} {\bibfnamefont {N.}~\bibnamefont
  {Bruot}}\ and\ \bibinfo {author} {\bibfnamefont {P.}~\bibnamefont {Cicuta}},\
  }\href {https://doi.org/10.1146/annurev-conmatphys-031115-011451} {\bibfield
  {journal} {\bibinfo  {journal} {Annu. Rev. Condens. Matter Phys.}\ }\textbf
  {\bibinfo {volume} {7}},\ \bibinfo {pages} {323} (\bibinfo {year}
  {2016})}\BibitemShut {NoStop}%
\bibitem [{\citenamefont {Brumley}\ \emph {et~al.}(2016)\citenamefont
  {Brumley}, \citenamefont {Bruot}, \citenamefont {Kotar}, \citenamefont
  {Goldstein}, \citenamefont {Cicuta},\ and\ \citenamefont {Polin}}]{Brumley}%
  \BibitemOpen
  \bibfield  {author} {\bibinfo {author} {\bibfnamefont {D.~R.}\ \bibnamefont
  {Brumley}}, \bibinfo {author} {\bibfnamefont {N.}~\bibnamefont {Bruot}},
  \bibinfo {author} {\bibfnamefont {J.}~\bibnamefont {Kotar}}, \bibinfo
  {author} {\bibfnamefont {R.~E.}\ \bibnamefont {Goldstein}}, \bibinfo {author}
  {\bibfnamefont {P.}~\bibnamefont {Cicuta}},\ and\ \bibinfo {author}
  {\bibfnamefont {M.}~\bibnamefont {Polin}},\ }\href
  {https://doi.org/10.1103/PhysRevFluids.1.081201} {\bibfield  {journal}
  {\bibinfo  {journal} {Phys. Rev. Fluids}\ }\textbf {\bibinfo {volume} {1}},\
  \bibinfo {pages} {081201} (\bibinfo {year} {2016})}\BibitemShut {NoStop}%
\bibitem [{\citenamefont {Lagomarsino}\ \emph {et~al.}(2003)\citenamefont
  {Lagomarsino}, \citenamefont {Jona},\ and\ \citenamefont
  {Bassetti}}]{Consentino}%
  \BibitemOpen
  \bibfield  {author} {\bibinfo {author} {\bibfnamefont {M.~C.}\ \bibnamefont
  {Lagomarsino}}, \bibinfo {author} {\bibfnamefont {P.}~\bibnamefont {Jona}},\
  and\ \bibinfo {author} {\bibfnamefont {B.}~\bibnamefont {Bassetti}},\ }\href
  {https://doi.org/10.1103/PhysRevE.68.021908} {\bibfield  {journal} {\bibinfo
  {journal} {Phys. Rev. E}\ }\textbf {\bibinfo {volume} {68}},\ \bibinfo
  {pages} {021908} (\bibinfo {year} {2003})}\BibitemShut {NoStop}%
\bibitem [{\citenamefont {Bruot}\ \emph {et~al.}(2012)\citenamefont {Bruot},
  \citenamefont {Kotar}, \citenamefont {de~Lillo}, \citenamefont
  {Cosentino~Lagomarsino},\ and\ \citenamefont {Cicuta}}]{Bruot}%
  \BibitemOpen
  \bibfield  {author} {\bibinfo {author} {\bibfnamefont {N.}~\bibnamefont
  {Bruot}}, \bibinfo {author} {\bibfnamefont {J.}~\bibnamefont {Kotar}},
  \bibinfo {author} {\bibfnamefont {F.}~\bibnamefont {de~Lillo}}, \bibinfo
  {author} {\bibfnamefont {M.}~\bibnamefont {Cosentino~Lagomarsino}},\ and\
  \bibinfo {author} {\bibfnamefont {P.}~\bibnamefont {Cicuta}},\ }\href
  {https://doi.org/10.1103/PhysRevLett.109.164103} {\bibfield  {journal}
  {\bibinfo  {journal} {Phys. Rev. Lett.}\ }\textbf {\bibinfo {volume} {109}},\
  \bibinfo {pages} {164103} (\bibinfo {year} {2012})}\BibitemShut {NoStop}%
\bibitem [{\citenamefont {Niedermayer}\ \emph {et~al.}(2008)\citenamefont
  {Niedermayer}, \citenamefont {Eckhardt},\ and\ \citenamefont
  {Lenz}}]{NiederMayer}%
  \BibitemOpen
  \bibfield  {author} {\bibinfo {author} {\bibfnamefont {T.}~\bibnamefont
  {Niedermayer}}, \bibinfo {author} {\bibfnamefont {B.}~\bibnamefont
  {Eckhardt}},\ and\ \bibinfo {author} {\bibfnamefont {P.}~\bibnamefont
  {Lenz}},\ }\href {https://doi.org/10.1063/1.2956984} {\bibfield  {journal}
  {\bibinfo  {journal} {Chaos}\ }\textbf {\bibinfo {volume} {18}},\ \bibinfo
  {pages} {037128} (\bibinfo {year} {2008})}\BibitemShut {NoStop}%
\bibitem [{\citenamefont {Uchida}\ and\ \citenamefont
  {Golestanian}(2011)}]{Uchida2011}%
  \BibitemOpen
  \bibfield  {author} {\bibinfo {author} {\bibfnamefont {N.}~\bibnamefont
  {Uchida}}\ and\ \bibinfo {author} {\bibfnamefont {R.}~\bibnamefont
  {Golestanian}},\ }\href {https://doi.org/10.1103/PhysRevLett.106.058104}
  {\bibfield  {journal} {\bibinfo  {journal} {Phys. Rev. Lett.}\ }\textbf
  {\bibinfo {volume} {106}},\ \bibinfo {pages} {058104} (\bibinfo {year}
  {2011})}\BibitemShut {NoStop}%
\bibitem [{\citenamefont {Uchida}\ and\ \citenamefont
  {Golestanian}(2012)}]{Uchida2012}%
  \BibitemOpen
  \bibfield  {author} {\bibinfo {author} {\bibfnamefont {N.}~\bibnamefont
  {Uchida}}\ and\ \bibinfo {author} {\bibfnamefont {R.}~\bibnamefont
  {Golestanian}},\ }\href {https://doi.org/10.1140/epje/i2012-12135-5}
  {\bibfield  {journal} {\bibinfo  {journal} {Eur. Phys. J. E}\ }\textbf
  {\bibinfo {volume} {35}},\ \bibinfo {pages} {135} (\bibinfo {year}
  {2012})}\BibitemShut {NoStop}%
\bibitem [{\citenamefont {Uchida}\ \emph {et~al.}(2017)\citenamefont {Uchida},
  \citenamefont {Golestanian},\ and\ \citenamefont {Bennett}}]{Uchida2017}%
  \BibitemOpen
  \bibfield  {author} {\bibinfo {author} {\bibfnamefont {N.}~\bibnamefont
  {Uchida}}, \bibinfo {author} {\bibfnamefont {R.}~\bibnamefont
  {Golestanian}},\ and\ \bibinfo {author} {\bibfnamefont {R.~R.}\ \bibnamefont
  {Bennett}},\ }\href {https://doi.org/10.7566/JPSJ.86.101007} {\bibfield
  {journal} {\bibinfo  {journal} {J. Phys. Soc. Jpn.}\ }\textbf {\bibinfo
  {volume} {86}},\ \bibinfo {pages} {101007} (\bibinfo {year}
  {2017})}\BibitemShut {NoStop}%
\bibitem [{\citenamefont {Kotar}\ \emph {et~al.}(2013)\citenamefont {Kotar},
  \citenamefont {Debono}, \citenamefont {Bruot}, \citenamefont {Box},
  \citenamefont {Phillips}, \citenamefont {Simpson}, \citenamefont {Hanna},\
  and\ \citenamefont {Cicuta}}]{Kotar}%
  \BibitemOpen
  \bibfield  {author} {\bibinfo {author} {\bibfnamefont {J.}~\bibnamefont
  {Kotar}}, \bibinfo {author} {\bibfnamefont {L.}~\bibnamefont {Debono}},
  \bibinfo {author} {\bibfnamefont {N.}~\bibnamefont {Bruot}}, \bibinfo
  {author} {\bibfnamefont {S.}~\bibnamefont {Box}}, \bibinfo {author}
  {\bibfnamefont {D.}~\bibnamefont {Phillips}}, \bibinfo {author}
  {\bibfnamefont {S.}~\bibnamefont {Simpson}}, \bibinfo {author} {\bibfnamefont
  {S.}~\bibnamefont {Hanna}},\ and\ \bibinfo {author} {\bibfnamefont
  {P.}~\bibnamefont {Cicuta}},\ }\href
  {https://doi.org/10.1103/PhysRevLett.111.228103} {\bibfield  {journal}
  {\bibinfo  {journal} {Phys. Rev. Lett.}\ }\textbf {\bibinfo {volume} {111}},\
  \bibinfo {pages} {228103} (\bibinfo {year} {2013})}\BibitemShut {NoStop}%
\bibitem [{\citenamefont {Theers}\ and\ \citenamefont {Winkler}(2013)}]{Mario}%
  \BibitemOpen
  \bibfield  {author} {\bibinfo {author} {\bibfnamefont {M.}~\bibnamefont
  {Theers}}\ and\ \bibinfo {author} {\bibfnamefont {R.~G.}\ \bibnamefont
  {Winkler}},\ }\href {https://doi.org/10.1103/PhysRevE.88.023012} {\bibfield
  {journal} {\bibinfo  {journal} {Phys. Rev. E}\ }\textbf {\bibinfo {volume}
  {88}},\ \bibinfo {pages} {023012} (\bibinfo {year} {2013})}\BibitemShut
  {NoStop}%
\bibitem [{\citenamefont {Oyama}\ \emph {et~al.}(2018)\citenamefont {Oyama},
  \citenamefont {Teshigawara}, \citenamefont {Molina}, \citenamefont
  {Yamamoto},\ and\ \citenamefont {Taniguchi}}]{Oyama2018}%
  \BibitemOpen
  \bibfield  {author} {\bibinfo {author} {\bibfnamefont {N.}~\bibnamefont
  {Oyama}}, \bibinfo {author} {\bibfnamefont {K.}~\bibnamefont {Teshigawara}},
  \bibinfo {author} {\bibfnamefont {J.~J.}\ \bibnamefont {Molina}}, \bibinfo
  {author} {\bibfnamefont {R.}~\bibnamefont {Yamamoto}},\ and\ \bibinfo
  {author} {\bibfnamefont {T.}~\bibnamefont {Taniguchi}},\ }\href
  {https://doi.org/10.1103/PhysRevE.97.032611} {\bibfield  {journal} {\bibinfo
  {journal} {Phys. Rev. E}\ }\textbf {\bibinfo {volume} {97}},\ \bibinfo
  {pages} {032611} (\bibinfo {year} {2018})},\ \BibitemShut {NoStop}%
\bibitem [{\citenamefont {Roichman}\ \emph {et~al.}(2007)\citenamefont
  {Roichman}, \citenamefont {Grier},\ and\ \citenamefont
  {Zaslavsky}}]{Roichman}%
  \BibitemOpen
  \bibfield  {author} {\bibinfo {author} {\bibfnamefont {Y.}~\bibnamefont
  {Roichman}}, \bibinfo {author} {\bibfnamefont {D.~G.}\ \bibnamefont
  {Grier}},\ and\ \bibinfo {author} {\bibfnamefont {G.}~\bibnamefont
  {Zaslavsky}},\ }\href {https://doi.org/10.1103/PhysRevE.75.020401(R)} {\bibfield
   {journal} {\bibinfo  {journal} {Phys. Rev. E}\ }\textbf {\bibinfo {volume}
  {75}},\ \bibinfo {pages} {020401(R)} (\bibinfo {year} {2007})}\BibitemShut
  {NoStop}%
\bibitem [{\citenamefont {Sokolov}\ \emph {et~al.}(2011)\citenamefont
  {Sokolov}, \citenamefont {Frydel}, \citenamefont {Grier}, \citenamefont
  {Diamant},\ and\ \citenamefont {Roichman}}]{Sokolov}%
  \BibitemOpen
  \bibfield  {author} {\bibinfo {author} {\bibfnamefont {Y.}~\bibnamefont
  {Sokolov}}, \bibinfo {author} {\bibfnamefont {D.}~\bibnamefont {Frydel}},
  \bibinfo {author} {\bibfnamefont {D.~G.}\ \bibnamefont {Grier}}, \bibinfo
  {author} {\bibfnamefont {H.}~\bibnamefont {Diamant}},\ and\ \bibinfo {author}
  {\bibfnamefont {Y.}~\bibnamefont {Roichman}},\ }\href
  {https://doi.org/10.1103/PhysRevLett.107.158302} {\bibfield  {journal}
  {\bibinfo  {journal} {Phys. Rev. Lett.}\ }\textbf {\bibinfo {volume} {107}},\
  \bibinfo {pages} {158302} (\bibinfo {year} {2011})}\BibitemShut {NoStop}%
\bibitem [{\citenamefont {Sassa}\ \emph {et~al.}(2012)\citenamefont {Sassa},
  \citenamefont {Shibata}, \citenamefont {Iwashita},\ and\ \citenamefont
  {Kimura}}]{Sassa}%
  \BibitemOpen
  \bibfield  {author} {\bibinfo {author} {\bibfnamefont {Y.}~\bibnamefont
  {Sassa}}, \bibinfo {author} {\bibfnamefont {S.}~\bibnamefont {Shibata}},
  \bibinfo {author} {\bibfnamefont {Y.}~\bibnamefont {Iwashita}},\ and\
  \bibinfo {author} {\bibfnamefont {Y.}~\bibnamefont {Kimura}},\ }\href
  {https://doi.org/10.1103/PhysRevE.85.061402} {\bibfield  {journal} {\bibinfo
  {journal} {Phys. Rev. E}\ }\textbf {\bibinfo {volume} {85}},\ \bibinfo
  {pages} {061402} (\bibinfo {year} {2012})}\BibitemShut {NoStop}%
\bibitem [{\citenamefont {Okubo}\ \emph {et~al.}(2015)\citenamefont {Okubo},
  \citenamefont {Shibata}, \citenamefont {Kawamura}, \citenamefont {Ichikawa},\
  and\ \citenamefont {Kimura}}]{Okubo2015}%
  \BibitemOpen
  \bibfield  {author} {\bibinfo {author} {\bibfnamefont {S.}~\bibnamefont
  {Okubo}}, \bibinfo {author} {\bibfnamefont {S.}~\bibnamefont {Shibata}},
  \bibinfo {author} {\bibfnamefont {Y.~S.}\ \bibnamefont {Kawamura}}, \bibinfo
  {author} {\bibfnamefont {M.}~\bibnamefont {Ichikawa}},\ and\ \bibinfo
  {author} {\bibfnamefont {Y.}~\bibnamefont {Kimura}},\ }\href
  {https://doi.org/10.1103/PhysRevE.92.032303} {\bibfield  {journal} {\bibinfo
  {journal} {Phys. Rev. E}\ }\textbf {\bibinfo {volume} {92}},\ \bibinfo
  {pages} {032303} (\bibinfo {year} {2015})}\BibitemShut {NoStop}%
\bibitem [{\citenamefont {Kimura}(2017)}]{Kimura2017}%
  \BibitemOpen
  \bibfield  {author} {\bibinfo {author} {\bibfnamefont {Y.}~\bibnamefont
  {Kimura}},\ }\href {https://doi.org/10.7566/JPSJ.86.101003} {\bibfield
  {journal} {\bibinfo  {journal} {J. Phys. Soc. Jpn.}\ }\textbf {\bibinfo
  {volume} {86}},\ \bibinfo {pages} {101003} (\bibinfo {year}
  {2017})}\BibitemShut {NoStop}%
\bibitem [{\citenamefont {Nagar}\ and\ \citenamefont {Roichman}(2014)}]{Nagar}%
  \BibitemOpen
  \bibfield  {author} {\bibinfo {author} {\bibfnamefont {H.}~\bibnamefont
  {Nagar}}\ and\ \bibinfo {author} {\bibfnamefont {Y.}~\bibnamefont
  {Roichman}},\ }\href {https://doi.org/10.1103/PhysRevE.90.042302} {\bibfield
  {journal} {\bibinfo  {journal} {Phys. Rev. E}\ }\textbf {\bibinfo {volume}
  {90}},\ \bibinfo {pages} {042302} (\bibinfo {year} {2014})}\BibitemShut
  {NoStop}%
\bibitem [{\citenamefont {Lutz}\ \emph {et~al.}(2006)\citenamefont {Lutz},
  \citenamefont {Reichert}, \citenamefont {Stark},\ and\ \citenamefont
  {Bechinger}}]{Lutz}%
  \BibitemOpen
  \bibfield  {author} {\bibinfo {author} {\bibfnamefont {C.}~\bibnamefont
  {Lutz}}, \bibinfo {author} {\bibfnamefont {M.}~\bibnamefont {Reichert}},
  \bibinfo {author} {\bibfnamefont {H.}~\bibnamefont {Stark}},\ and\ \bibinfo
  {author} {\bibfnamefont {C.}~\bibnamefont {Bechinger}},\ }\href
  {https://doi.org/10.1209/epl/i2006-10017-9} {\bibfield  {journal} {\bibinfo
  {journal} {Europhys. Lett.}\ }\textbf {\bibinfo {volume} {74}},\ \bibinfo
  {pages} {719} (\bibinfo {year} {2006})}\BibitemShut {NoStop}%
\bibitem [{\citenamefont {Kotar}\ \emph {et~al.}(2010)\citenamefont {Kotar},
  \citenamefont {Leoni}, \citenamefont {Bassetti}, \citenamefont
  {Lagomarsino},\ and\ \citenamefont {Cicuta}}]{Kotar2010}%
  \BibitemOpen
  \bibfield  {author} {\bibinfo {author} {\bibfnamefont {J.}~\bibnamefont
  {Kotar}}, \bibinfo {author} {\bibfnamefont {M.}~\bibnamefont {Leoni}},
  \bibinfo {author} {\bibfnamefont {B.}~\bibnamefont {Bassetti}}, \bibinfo
  {author} {\bibfnamefont {M.~C.}\ \bibnamefont {Lagomarsino}},\ and\ \bibinfo
  {author} {\bibfnamefont {P.}~\bibnamefont {Cicuta}},\ }\href
  {https://doi.org/10.1073/pnas.0912455107} {\bibfield  {journal} {\bibinfo
  {journal} {Proc. Natl. Acad. Sci.}\ }\textbf {\bibinfo {volume} {107}},\
  \bibinfo {pages} {7669} (\bibinfo {year} {2010})}\BibitemShut {NoStop}%
\bibitem [{\citenamefont {Lhermerout}\ \emph {et~al.}(2012)\citenamefont
  {Lhermerout}, \citenamefont {Bruot}, \citenamefont {Cicuta}, \citenamefont
  {Kotar},\ and\ \citenamefont {Cicuta}}]{Lhermerout_2012}%
  \BibitemOpen
  \bibfield  {author} {\bibinfo {author} {\bibfnamefont {R.}~\bibnamefont
  {Lhermerout}}, \bibinfo {author} {\bibfnamefont {N.}~\bibnamefont {Bruot}},
  \bibinfo {author} {\bibfnamefont {G.~M.}\ \bibnamefont {Cicuta}}, \bibinfo
  {author} {\bibfnamefont {J.}~\bibnamefont {Kotar}},\ and\ \bibinfo {author}
  {\bibfnamefont {P.}~\bibnamefont {Cicuta}},\ }\href
  {https://doi.org/10.1088/1367-2630/14/10/105023} {\bibfield  {journal}
  {\bibinfo  {journal} {New J. Phys.}\ }\textbf {\bibinfo {volume} {14}},\
  \bibinfo {pages} {105023} (\bibinfo {year} {2012})}\BibitemShut {NoStop}%
\bibitem [{\citenamefont {Kavre}\ \emph {et~al.}(2015)\citenamefont {Kavre},
  \citenamefont {Vilfan},\ and\ \citenamefont {Babi\ifmmode~\check{c}\else
  \v{c}\fi{}}}]{Kavre}%
  \BibitemOpen
  \bibfield  {author} {\bibinfo {author} {\bibfnamefont {I.}~\bibnamefont
  {Kavre}}, \bibinfo {author} {\bibfnamefont {A.}~\bibnamefont {Vilfan}},\ and\
  \bibinfo {author} {\bibfnamefont {D.~c.~v.}\ \bibnamefont
  {Babi\ifmmode~\check{c}\else \v{c}\fi{}}},\ }\href
  {https://doi.org/10.1103/PhysRevE.91.031002(R)} {\bibfield  {journal} {\bibinfo
  {journal} {Phys. Rev. E}\ }\textbf {\bibinfo {volume} {91}},\ \bibinfo
  {pages} {031002(R)} (\bibinfo {year} {2015})}\BibitemShut {NoStop}%
\bibitem [{\citenamefont {Simpson}\ \emph {et~al.}(2016)\citenamefont
  {Simpson}, \citenamefont {Chv\'atal},\ and\ \citenamefont
  {Zem\'anek}}]{Simpson}%
  \BibitemOpen
  \bibfield  {author} {\bibinfo {author} {\bibfnamefont {S.~H.}\ \bibnamefont
  {Simpson}}, \bibinfo {author} {\bibfnamefont {L.}~\bibnamefont {Chv\'atal}},\
  and\ \bibinfo {author} {\bibfnamefont {P.}~\bibnamefont {Zem\'anek}},\ }\href
  {https://doi.org/10.1103/PhysRevA.93.023842} {\bibfield  {journal} {\bibinfo
  {journal} {Phys. Rev. A}\ }\textbf {\bibinfo {volume} {93}},\ \bibinfo
  {pages} {023842} (\bibinfo {year} {2016})}\BibitemShut {NoStop}%
\bibitem [{\citenamefont {Di~Leonardo}\ \emph {et~al.}(2012)\citenamefont
  {Di~Leonardo}, \citenamefont {B\'uz\'as}, \citenamefont {Kelemen},
  \citenamefont {Vizsnyiczai}, \citenamefont {Oroszi},\ and\ \citenamefont
  {Ormos}}]{Leonardo}%
  \BibitemOpen
  \bibfield  {author} {\bibinfo {author} {\bibfnamefont {R.}~\bibnamefont
  {Di~Leonardo}}, \bibinfo {author} {\bibfnamefont {A.}~\bibnamefont
  {B\'uz\'as}}, \bibinfo {author} {\bibfnamefont {L.}~\bibnamefont {Kelemen}},
  \bibinfo {author} {\bibfnamefont {G.}~\bibnamefont {Vizsnyiczai}}, \bibinfo
  {author} {\bibfnamefont {L.}~\bibnamefont {Oroszi}},\ and\ \bibinfo {author}
  {\bibfnamefont {P.}~\bibnamefont {Ormos}},\ }\href
  {https://doi.org/10.1103/PhysRevLett.109.034104} {\bibfield  {journal}
  {\bibinfo  {journal} {Phys. Rev. Lett.}\ }\textbf {\bibinfo {volume} {109}},\
  \bibinfo {pages} {034104} (\bibinfo {year} {2012})}\BibitemShut {NoStop}%
\bibitem [{\citenamefont {Curtis}\ and\ \citenamefont {Grier}(2003)}]{Curtis}%
  \BibitemOpen
  \bibfield  {author} {\bibinfo {author} {\bibfnamefont {J.~E.}\ \bibnamefont
  {Curtis}}\ and\ \bibinfo {author} {\bibfnamefont {D.~G.}\ \bibnamefont
  {Grier}},\ }\href {https://doi.org/10.1103/PhysRevLett.90.133901} {\bibfield
  {journal} {\bibinfo  {journal} {Phys. Rev. Lett.}\ }\textbf {\bibinfo
  {volume} {90}},\ \bibinfo {pages} {133901} (\bibinfo {year}
  {2003})}\BibitemShut {NoStop}%
\bibitem [{\citenamefont {Reichert}\ and\ \citenamefont
  {Stark}(2004)}]{Reichert2004}%
  \BibitemOpen
  \bibfield  {author} {\bibinfo {author} {\bibfnamefont {M.}~\bibnamefont
  {Reichert}}\ and\ \bibinfo {author} {\bibfnamefont {H.}~\bibnamefont
  {Stark}},\ }\href {https://doi.org/10.1088/0953-8984/16/38/023} {\bibfield
  {journal} {\bibinfo  {journal} {J. Phys. Condens. Matter}\ }\textbf {\bibinfo
  {volume} {16}},\ \bibinfo {pages} {S4085} (\bibinfo {year}
  {2004})}\BibitemShut {NoStop}%
\bibitem [{\citenamefont {Saito}\ \emph {et~al.}(2018)\citenamefont {Saito},
  \citenamefont {Okubo},\ and\ \citenamefont {Kimura}}]{Saito2018}%
  \BibitemOpen
  \bibfield  {author} {\bibinfo {author} {\bibfnamefont {K.}~\bibnamefont
  {Saito}}, \bibinfo {author} {\bibfnamefont {S.}~\bibnamefont {Okubo}},\ and\
  \bibinfo {author} {\bibfnamefont {Y.}~\bibnamefont {Kimura}},\ }\href
  {https://doi.org/10.1039/C8SM00582F} {\bibfield  {journal} {\bibinfo
  {journal} {Soft Matter}\ }\textbf {\bibinfo {volume} {14}},\ \bibinfo {pages}
  {6037} (\bibinfo {year} {2018})}\BibitemShut {NoStop}%
\bibitem [{\citenamefont {Rotne}\ and\ \citenamefont
  {Prager}(1969)}]{Rotne1969}%
  \BibitemOpen
  \bibfield  {author} {\bibinfo {author} {\bibfnamefont {J.}~\bibnamefont
  {Rotne}}\ and\ \bibinfo {author} {\bibfnamefont {S.}~\bibnamefont {Prager}},\
  }\href {https://doi.org/10.1063/1.1670977} {\bibfield  {journal} {\bibinfo
  {journal} {J. Chem. Phys.}\ }\textbf {\bibinfo {volume} {50}},\ \bibinfo
  {pages} {4831} (\bibinfo {year} {1969})}\BibitemShut {NoStop}%
\bibitem [{\citenamefont {Yamakawa}(1970)}]{Yamakawa}%
  \BibitemOpen
  \bibfield  {author} {\bibinfo {author} {\bibfnamefont {H.}~\bibnamefont
  {Yamakawa}},\ }\href {https://doi.org/10.1063/1.1673799} {\bibfield
  {journal} {\bibinfo  {journal} {J. Chem. Phys.}\ }\textbf {\bibinfo {volume}
  {53}},\ \bibinfo {pages} {436} (\bibinfo {year} {1970})}\BibitemShut
  {NoStop}%
\bibitem [{\citenamefont {Kim}\ and\ \citenamefont {Karrila}(1991)}]{Kim}%
  \BibitemOpen
  \bibfield  {author} {\bibinfo {author} {\bibfnamefont {S.}~\bibnamefont
  {Kim}}\ and\ \bibinfo {author} {\bibfnamefont {S.~J.}\ \bibnamefont
  {Karrila}},\ }\href@noop {} {\emph {\bibinfo {title}
  {Microhydrodynamics-Principles and Selected Applications}}}\ (\bibinfo
  {publisher} {Butterworth-Heinemann},\ \bibinfo {year} {1991})\BibitemShut
  {NoStop}%
\end{thebibliography}

%

\end{document}